\documentclass[12pt,onecolumn,draftcls]{IEEEtran}
\usepackage{verbatim}
\usepackage{graphicx}
\usepackage{amsmath,amssymb,amsfonts,bm}
\usepackage{subfigure}
\usepackage{psfrag}
\usepackage[dvips]{epsfig}
\usepackage{amsthm}
\usepackage{color}
\usepackage[usenames,dvipsnames]{pstricks}
\usepackage[dvips]{epsfig}
\usepackage{pst-grad} 
\usepackage{pst-plot} 
\usepackage{tikz,pgf}
\usetikzlibrary{decorations.text}
\usepackage{amsmath,graphicx}
\usepackage{amsmath,graphicx}
\usepackage{enumerate}
\usepackage{amsbsy}
\usepackage{amssymb}
\usepackage{amsthm}
\usepackage{amscd}
\usepackage{subfigure}
\usepackage{color}
\usepackage{cite}
\usepackage{listings}

\usepackage{tikz}
\usepackage{pgf}

\usepackage{MnSymbol}
\usepackage{stackrel}

\newtheorem{rema}{Remark}

\def\bA{\mathbf{A}}

\def\bB{\mathbf{B}}

\def\bI{\mathbf{I}}

\def\bone\mathbf{1}
\def\bone{{\boldsymbol 1}}

\def\bT{\mathbf{T}}

\def\bv{\mathbf{v}}
\def\bv{\mathbf{v}}

\def\bx{\mathbf{x}}
\def\bx{\mathbf{x}}

\def\bz{\mathbf{z}}

\def\chx{18-2}
\def\chxa{23-2}
\def\chxad{\chxa+14}
\def\chxas{\chxa}
\def\chxb{27-2}
\def\chxbd{\chxb+14}
\def\chxbs{\chxb}
\def\chxc{23-2}
\def\chxcd{\chxc+14}
\def\chxcs{\chxc}
\def\chxd{\chx+14}
\def\chxs{\chx}
\def\chy{11}
\def\chya{15}
\def\chyad{\chya-8}
\def\chyas{\chyad-8}
\def\chyb{11}
\def\chybd{\chyb-8}
\def\chybs{\chybd-8}
\def\chyc{7}
\def\chycd{\chyc-8}
\def\chycs{\chycd-8}
\def\chyd{\chy-8}
\def\chys{\chyd-8}
\def\complexC{{\mathbb{C}}}

\begin{document}

\title{\LARGE Cooperative Fair Throughput Maximization in a Multi-Cluster Wireless Powered 
Network}

\author{Omid Rezaei, Maryam Masjedi, Mohammad Mahdi Naghsh\IEEEauthorrefmark{1}, Saeed Gazor, and Mohammad Mahdi Nayebi

\thanks{O. Rezaei and M. M. Nayebi are with the Department of Electrical Engineering, Sharif University of Technology, Tehran, 11155-4363, Iran. M. Masjedi, and M. M. Naghsh are with the Department of Electrical and Computer Engineering, Isfahan University of Technology, Isfahan, 84156-83111, Iran. S. Gazor is with the Department of Electrical and Computer Engineering, Queen’s University, Kingston, Ontario, K7L 3N6, Canada.
	
\IEEEauthorrefmark{1}Please address all the correspondence to M. M. Naghsh, Phone: (+98) 31-33912450; Fax: (+98) 31-33912451; Email: mm\_naghsh@cc.iut.ac.ir}
}
\maketitle
\begin{abstract}
In this paper, we study a multi-cluster Wireless Powered Communication Network (WPCN) where each cluster of users cooperate with a Cluster Head (CH) and a Hybrid Access Point (HAP). All users are equipped with multiple antennas. This HAP employs beamforming to deliver energy to the users in the downlink phase. The users of each cluster transmit their signals to the HAP and to their CHs in the uplink phase. In the next step the CHs first relay the signals of their cluster users and then transmit their own information to the HAP. We design the Energy Beamforming (EB) matrix, transmit covariance matrices of the users and allocate time slots to energy transfer and cooperation phases by optimizing the max-min and sum throughputs of the network. Our optimization is a non-convex problem which we break it to two non-convex sub-problems and solve them by employing the Alternating Optimization (AO) and the Minorization-Maximization (MM) technique. We recast the resulting sub-problems as a convex Second Order Cone Programming (SOCP) and a Quadratic Constraint Quadratic Programming (QCQP) for the max-min and sum throughput maximization problems, respectively. We take into account the consider imperfect Channel State Information (CSI) and non-linearity in Energy Harvesting (EH) circuits. Using numerical examples, we explore the impact of the proposed cooperation as well as optimal placement of CHs under various setups.
\end{abstract}
\pagebreak
\begin{IEEEkeywords}
Cluster-based communication, minorization-maximization, user cooperation, user fairness, wireless powered communication.
\end{IEEEkeywords}
\section{Introduction}
The operational lifetime of a communication network employing battery powered wireless devices can be enhanced by employing recent advances in Wireless Power Transfer (WPT) and harvesting technologies in which the wireless devices receive power/energy from dedicated wireless power transmitters \cite{ho2012optimal}. One important application of WPT is in Wireless Powered Communication Network (WPCN), where the wireless devices harvest the wireless energy during the downlink communication in order to power the uplink phase. In a WPCN, the energy transmitter and the information receiver are placed either at the same location as a Hybrid Access Point (HAP) or at separate locations \cite{bi2016placement}. The authors of \cite{Throughput} have proposed a Harvest-Then-Transmit (HTT) protocol for a WPCN consisting of ${K}$ single-antenna users and a single-antenna HAP. They have shown  that in a WPCN which employs a HAP, an inherent unfair ``doubly near-far" phenomenon causes the users which are far from the HAP to harvest less energy in the downlink where they require more energy to transmit their information in the uplink. Several methods have been proposed in the literature to overcome the doubly near-far phenomenon \cite{ju2014user,song2019sum,zheng2020intelligent,zhong2017user, gu2015adaptive,chen2015harvest,chen2014wireless,di2017optimal, chu2017energy, liang2017optimization ,zhong2015optimum,huang2016relaying,yuan2017multi,yuan2020optimizing,mao2020fairness,yuan2022interactive,na2020clustered}.

In \cite{ju2014user,song2019sum,zheng2020intelligent,zhong2017user, gu2015adaptive,chen2015harvest,chen2014wireless}, the user cooperation is proposed for a three-node WPCN assuming that all nodes are equipped with single-antenna. In the proposed methods in \cite{ju2014user} and \cite{song2019sum}, the closer user to the HAP transmits the signal of the more distant user along with its own signal to the HAP and the sum throughput of the network is maximized to find optimal time and power allocations. The method in \cite{ju2014user} is extended in \cite{zheng2020intelligent} for a system employing an Intelligent Reflecting Surface (IRS). In \cite{zhong2017user}, the max-min throughput fairness was considered for a single-pair WPCN, where both users harvest the transmitted energy from an energy source and then help each other to send their information to a destination node. The outage performance is studied for a single-user WPCN in \cite{gu2015adaptive} and \cite{chen2015harvest}, where the user sent its information to the HAP via a dedicated relay. A single-pair is considered in \cite{chen2014wireless} with cooperation in both downlink and uplink phases.

Several references employ multiple antennas energy transmitter and design the Energy Beamforming (EB) to focus the transmitted energy toward the specific receivers \cite{di2017optimal, chu2017energy, liang2017optimization ,zhong2015optimum,huang2016relaying}.
The throughput maximization is investigated for a WPCN in \cite{di2017optimal,chu2017energy,liang2017optimization} consisting of two cooperating single-antenna users and a multiple antennas HAP. The throughput of a single-user WPCN is calculated in \cite{zhong2015optimum} employing a dedicated single-antenna relay, where a multiple antennas energy source first powers the single-antenna user and the relay. Then, the source transmits its information to a single-antenna destination node with the help of the relay. In \cite{huang2016relaying} a three-node cooperative wireless powered scenario is considered consisting of a source, destination and a dedicated relay which all have multiple antennas. The system in \cite{huang2016relaying} is designed to maximize the throughput.

The above works have considered the WPCN only in the context of a single-pair of users. This paradigm is extended in \cite{yuan2017multi,yuan2020optimizing,mao2020fairness} to consider multiple users in a single-cluster. In \cite{yuan2017multi}, all single-antenna users transmit their signals to the single-antenna cluster head (CH) and then the CH transmits the user signals as well as its own signal to the multiple antennas HAP via Time-Division-Multiple-Access (TDMA) protocol. The transmit time allocations and EB matrix were designed to improve the throughput fairness among the users. The model in \cite{yuan2017multi} is extended in \cite{yuan2020optimizing} to a cognitive WPCN by considering a secondary single-cluster WPCN and a primary point-to-point communication.
In \cite{mao2020fairness}, a fairness-aware transmission is employed for a single-cluster WPCN where each user sends its information to all the other users and cooperatively send their signals to the HAP using TDMA.

Grouping of users into clusters may practically lead to significant improvement in rate and energy \cite{ding2016general}. Similar order of enhancement is obtained in the throughput fairness for a two-cluster cooperating WPCN employing two multiple antennas HAPs and multiple single-antenna users in \cite{yuan2022interactive} where all the information transmissions in \cite{yuan2022interactive} are divided in time domain to avoid inter-cluster interference which increases the user latency. The sum throughput maximization is studied in \cite{na2020clustered} for a multi-cluster non-cooperating WPCN where HAP and all users have single-antenna.
To the best of our knowledge, the design of  multi-cluster cooperating WPCN with inter-cluster interference has not been addressed
in the literature. In this paper, we evaluate the minimum and sum throughput enhancement for a multi-cluster Multiple-Input Multiple-Output (MIMO) WPCN with user cooperation. We consider the general MIMO case and manage the inter-cluster interference where the main advantages of our proposed scheme are: (i) lower user latency in comparison with \cite{yuan2022interactive}; (ii) lower path loss attenuation contrary to \cite{na2020clustered}.
Precisely, the main contributions of this manuscript are summarized as follows:
\begin{itemize}
\item
We propose a new multi-cluster WPCN under HTT protocol with user cooperation, where the HAP as well as users have multiple antennas all operating in the same frequency band. As shown in Fig. \ref{ht}, the CH assists its cluster members to transmit their informations to the HAP more efficiently \cite{astaneh2009resource}. Moreover, in the downlink phase, the HAP employs beamforming technique to transfer energy to the users. In the uplink phase, the users in each cluster transmit their signals to the HAP and to their CH. Finally, the CHs first relay the signals of their cluster members and then transmit their own information signals to the HAP. Note that we must deal with the inter-cluster interference in this model.
\item
We design the EB matrix, transmit covariance matrices of the users and time allocations among energy transfer and cooperation phases in order to maximize max-min and sum throughputs of the network. These problems are non-convex and consequently are hard to solve. We propose a proper minorizer and develop algorithms (using Alternating Optimization (AO) and minorization-maximization (MM) techniques) to solve these problems. We recast the main sub-problems for max-min and sum throughput cases as Second Order Cone Programming (SOCP) and Quadratic Constraint Quadratic Programming (QCQP).
\item
We also investigate the impact of  imperfect Channel State Information (CSI) as well as the non-linearity in the Energy Harvesting (EH) circuits in our design problem and extend the proposed design to the case of deterministic energy signal.
\item
Via numerical examples, we explore the impact of cooperation and non-linear EH circuits as well as optimal location of the CHs.
\end{itemize}

\emph{Organization:}  We present the multi-cluster cooperative model for WPCN in Section~\ref{sys}. We analyze the max-min throughput optimization problem and present our proposed design method in Section~\ref{min}. We consider the special case of deterministic energy signal in Section~\ref{further} and investigate the sum throughput maximization problem. Numerical results in Section~\ref{num} reveal the efficiency of the proposed method. We finally draw conclusions in Section~\ref{con}.

\emph{Notation:} Bold lowercase (uppercase) letters are used for vectors (matrices).
The notations $\mathbb{E} [\cdot ]$, $\Re \lbrace \cdot \rbrace$, $\|\cdot\|_2$, ${(\cdot)^{{T}}}$, $(\cdot)^{{H}}$, $\mbox{tr} \lbrace \cdot \rbrace$, $\mathbf{\lambda}_{\textrm{max}}(\cdot)$, $\mathbf{\lambda}_{\textrm{min}}(\cdot)$, $\textrm{vec}(\cdot)$ and ${\nabla}^{2}_{{\mathbf{x}}} f(\cdot)$ indicate statistical expectation, real-part, $l_2$-norm of a vector, vector/matrix transpose, vector/matrix Hermitian, trace of a square matrix, principal eigenvalue of a Hermitian matrix, minimum eigenvalue of a Hermitian matrix, column-wise stacking of the elements of a matrix and the Hessian of the twice-differentiable function with respect to (w.r.t.) $\mathbf{x}$, respectively.
The symbol $\otimes$ stands for the Kronecker product of two matrices/vectors. We denote $\mathcal{CN}(\mathbf{\omega},\mathbf{\Sigma})$ as the circularly symmetric complex Gaussian (CSCG) distribution with mean $\mathbf{\omega}$ and covariance $\mathbf{\Sigma}$. The symbol $\mathbb{R}_+$ represents non-negative real numbers, $\complexC^{N \times N}$ ($\complexC^{N}$) is the set of ${N \times N}$ (${N \times 1}$) complex matrices (vectors), $\mathbb{S}_{++}^{N} \subseteq \complexC^{N \times N}$ denotes the set of positive definite matrices, and $\bI_{N}$ is the $N \times N$ identity matrix. Finally, the notation  $\bA \succ \bB$ ($\bA \succeq\bB$) means that $\bA-\bB$ is positive (semi)definite.
\section{System Model} \label{sys}
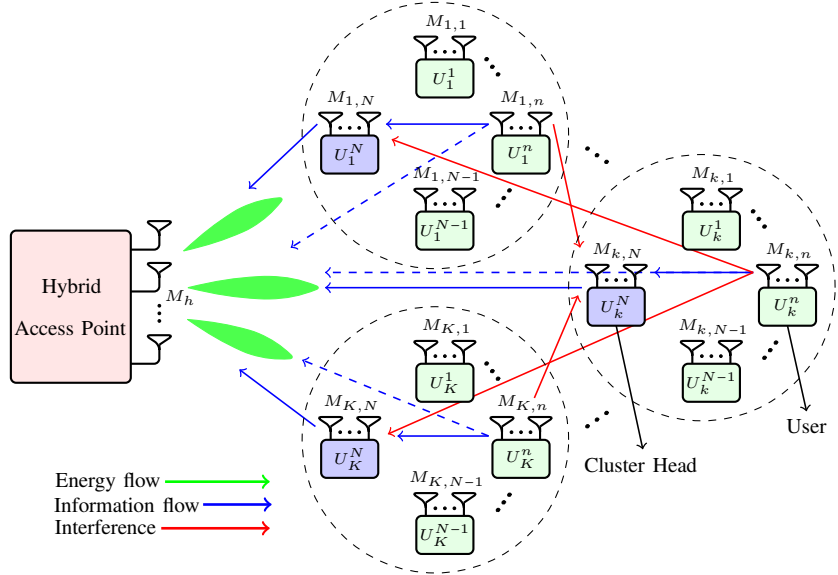
\begin{figure}
\centering

\begin{tikzpicture}[even odd rule,rounded corners=2pt,x=12pt,y=12pt,scale=.6]
\draw[thick,fill=red!10] (-.3,0) rectangle (6,8)
node[midway]{\scriptsize  \shortstack[l]{\scriptsize~~~ \hspace{-4pt} Hybrid \\  \\ \\ \hspace{5pt} \scriptsize \hspace{-11pt} Access Point}}; 
\draw[thick] (6,7)--++(1.5,0)--+(0,1);
\draw[thick] (7.5,8)--++(.75,0.5)--++(-1.5,0)--++(.75,-.5)--++(0,-.1);
\draw[thick] (6,4.8)--++(1.5,0)--+(0,1);
\draw[thick] (7.5,5.8)--++(.75,0.5)--++(-1.5,0)--++(.75,-.5)--++(0,-.1);
\draw[thick] (6,1)--++(1.5,0)--+(0,1);
\draw[thick] (7.5,2)--++(.75,0.5)--++(-1.5,0)--++(.75,-.5)--++(0,-.1);
\node [] at (8.6,4.5) {\tiny ${M}_h$};
\draw[fill=black!100] (7.5,4.05) circle (.07) ;
\draw[fill=black!100] (7.5,3.65) circle (.07) ;
\draw[fill=black!100] (7.5,3.25) circle (.07) ;

\draw [-,fill=green!40,green!70,rotate=28] (11,2) to [bend left=5] ++(3.5,.6) to [bend left=10]
++(2.5,-.6) to [bend left=10]
++(-2.5,-.6) to [bend left=5] ++(-3.5,.6);

\draw [-,fill=green!40,green!70] (9,5) to [bend left=5] ++(4,.6) to [bend left=10]
++(3,-.6) to [bend left=10]
++(-3,-.6) to [bend left=5] ++(-4,.6);

\draw [-,fill=green!40,green!70,rotate=-20] (7.3,6.2) to [bend left=5] ++(3.5,.6) to [bend left=10]
++(2.5,-.6) to [bend left=10]
++(-2.5,-.6) to [bend left=5] ++(-3.5,.6);
\draw[->,line width=.6,blue!100] (\chxbd-.3,\chybd+2.8)--+(-5.2,0);
\draw[->,dashed,line width=.6,blue!100] (\chxbd-.3,\chybd+2.8)--+(-22.5,0);
\draw[->,line width=.6,blue!100] (\chxd-.3,\chyd+2)--+(-13.5,0);

\draw[->,line width=.6,red!100] (\chxbd-.3,\chybd+2.8)--+(-19.2,-8.5);
\draw[->,line width=.6,red!100] (\chxbd-.3,\chybd+2.8)--+(-19,7);

\draw[->,line width=.6,red!100] (\chxb+3.2,\chyb+2.6)--+(1.4,-6.5);
\draw[->,line width=.6,blue!100] (\chxb-.2,\chyb+2.6)--+(-5.4,0);
\draw[->,dashed,line width=.6,blue!100] (\chxb-.2,\chyb+2.6)--+(-10.4,-6.5);
\draw[->,line width=.6,blue!100] (\chx-.2,\chy+2.6)--+(-3.6,-3.4);

\draw[->,line width=.6,red!100] (\chxbs+2.2,\chybs+4.2)--+(2.3,5.4);
\draw[->,line width=.6,blue!100] (\chxbs-.2,\chybs+2.2)--+(-4.8,0);
\draw[->,dashed,line width=.6,blue!100] (\chxbs-.2,\chybs+2.2)--+(-9.8,4);
\draw[->,line width=.6,blue!100] (\chxs-.3,\chys+2.7)--+(-4.,3);
\draw[thick,fill=blue!20] (\chx,\chy) rectangle ++(3,2) node[midway]{\tiny $U_{1}^{N}$};
\draw[thick] (\chx+.5,\chy+2)--+(0,.7);
\draw[thick] (\chx+.5,\chy+2.7)--++(.75,0.5)--++(-1.5,0)--++(.75,-.5)--++(0,-.1);
\draw[thick] (\chx+2.5,\chy+2)--+(0,.7);
\draw[thick] (\chx+2.5,\chy+2.7)--++(.75,0.5)--++(-1.5,0)--++(.75,-.5)--++(0,-.1);
\draw[fill=black!100] (\chx+1,\chy+2.4) circle (.07) ;
\draw[fill=black!100] (\chx+1.5,\chy+2.4) circle (.07) ;
\draw[fill=black!100] (\chx+2,\chy+2.4) circle (.07) ;
\node [] at (\chx+1.6,\chy+3.9) {\tiny ${M}_{1,N}$};

\draw[thick,fill=green!10] (\chxa,\chya) rectangle ++(3,2) node[midway]{\tiny $U_{1}^{1}$};
\draw[thick] (\chxa+.5,\chya+2)--+(0,.7);
\draw[thick] (\chxa+.5,\chya+2.7)--++(.75,0.5)--++(-1.5,0)--++(.75,-.5)--++(0,-.1);
\draw[thick] (\chxa+2.5,\chya+2)--+(0,.7);
\draw[thick] (\chxa+2.5,\chya+2.7)--++(.75,0.5)--++(-1.5,0)--++(.75,-.5)--++(0,-.1);
\draw[fill=black!100] (\chxa+1,\chya+2.4) circle (.07) ;
\draw[fill=black!100] (\chxa+1.5,\chya+2.4) circle (.07) ;
\draw[fill=black!100] (\chxa+2,\chya+2.4) circle (.07) ;
\node [] at (\chxa+1.6,\chya+3.9) {\tiny ${M}_{1,1}$};

\draw[thick,fill=green!10] (\chxb,\chyb) rectangle ++(3,2) node[midway]{\tiny $U_{1}^{n}$};
\draw[thick] (\chxb+.5,\chyb+2)--+(0,.7);
\draw[thick] (\chxb+.5,\chyb+2.7)--++(.75,0.5)--++(-1.5,0)--++(.75,-.5)--++(0,-.1);
\draw[thick] (\chxb+2.5,\chyb+2)--+(0,.7);
\draw[thick] (\chxb+2.5,\chyb+2.7)--++(.75,0.5)--++(-1.5,0)--++(.75,-.5)--++(0,-.1);
\draw[fill=black!100] (\chxb+1,\chyb+2.4) circle (.07) ;
\draw[fill=black!100] (\chxb+1.5,\chyb+2.4) circle (.07) ;
\draw[fill=black!100] (\chxb+2,\chyb+2.4) circle (.07) ;
\node [] at (\chxb+1.6,\chyb+3.9) {\tiny ${M}_{1,n}$};

\draw[thick,fill=green!10] (\chxc,\chyc) rectangle ++(3,2) node[midway]{\tiny $U_{1}^{N-1}$};
\draw[thick] (\chxc+.5,\chyc+2)--+(0,.7);
\draw[thick] (\chxc+.5,\chyc+2.7)--++(.75,0.5)--++(-1.5,0)--++(.75,-.5)--++(0,-.1);
\draw[thick] (\chxc+2.5,\chyc+2)--+(0,.7);
\draw[thick] (\chxc+2.5,\chyc+2.7)--++(.75,0.5)--++(-1.5,0)--++(.75,-.5)--++(0,-.1);
\draw[fill=black!100] (\chxc+1,\chyc+2.4) circle (.07) ;
\draw[fill=black!100] (\chxc+1.5,\chyc+2.4) circle (.07) ;
\draw[fill=black!100] (\chxc+2,\chyc+2.4) circle (.07) ;
\node [] at (\chxc+1.6,\chyc+3.9) {\tiny ${M}_{1,N-1}$};

\draw[dashed,fill=black!100] (\chxa+3.7,\chya+2) circle (.1) ;
\draw[dashed,fill=black!100] (\chxa+4,\chya+1.6) circle (.1) ;
\draw[dashed,fill=black!100] (\chxa+4.3,\chya+1.2) circle (.1) ;

\draw[dashed,fill=black!100] (\chxb+.9,\chyb-.9) circle (.1) ;
\draw[dashed,fill=black!100] (\chxb+.6,\chyb-1.3) circle (.1) ;
\draw[dashed,fill=black!100] (\chxb+.3,\chyb-1.7) circle (.1) ;

\draw[dashed,black!100] (\chx+6,\chy+2) circle (7) ;
\draw[thick,fill=blue!20] (\chxd,\chyd) rectangle ++(3,2) node[midway]{\tiny $U_{k}^{N}$};
\draw[thick] (\chxd+.5,\chyd+2)--+(0,.7);
\draw[thick] (\chxd+.5,\chyd+2.7)--++(.75,0.5)--++(-1.5,0)--++(.75,-.5)--++(0,-.1);
\draw[thick] (\chxd+2.5,\chyd+2)--+(0,.7);
\draw[thick] (\chxd+2.5,\chyd+2.7)--++(.75,0.5)--++(-1.5,0)--++(.75,-.5)--++(0,-.1);
\draw[fill=black!100] (\chxd+1,\chyd+2.4) circle (.07) ;
\draw[fill=black!100] (\chxd+1.5,\chyd+2.4) circle (.07) ;
\draw[fill=black!100] (\chxd+2,\chyd+2.4) circle (.07) ;
\node [] at (\chxd+1.6,\chyd+3.9) {\tiny ${M}_{k,N}$};

\draw[thick,fill=green!10] (\chxad,\chyad) rectangle ++(3,2) node[midway]{\tiny $U_{k}^{1}$};
\draw[thick] (\chxad+.5,\chyad+2)--+(0,.7);
\draw[thick] (\chxad+.5,\chyad+2.7)--++(.75,0.5)--++(-1.5,0)--++(.75,-.5)--++(0,-.1);
\draw[thick] (\chxad+2.5,\chyad+2)--+(0,.7);
\draw[thick] (\chxad+2.5,\chyad+2.7)--++(.75,0.5)--++(-1.5,0)--++(.75,-.5)--++(0,-.1);
\draw[fill=black!100] (\chxad+1,\chyad+2.4) circle (.07) ;
\draw[fill=black!100] (\chxad+1.5,\chyad+2.4) circle (.07) ;
\draw[fill=black!100] (\chxad+2,\chyad+2.4) circle (.07) ;
\node [] at (\chxad+1.6,\chyad+3.9) {\tiny ${M}_{k,1}$};

\draw[thick,fill=green!10] (\chxbd,\chybd) rectangle ++(3,2) node[midway]{\tiny $U_{k}^{n}$};
\draw[thick] (\chxbd+.5,\chybd+2)--+(0,.7);
\draw[thick] (\chxbd+.5,\chybd+2.7)--++(.75,0.5)--++(-1.5,0)--++(.75,-.5)--++(0,-.1);
\draw[thick] (\chxbd+2.5,\chybd+2)--+(0,.7);
\draw[thick] (\chxbd+2.5,\chybd+2.7)--++(.75,0.5)--++(-1.5,0)--++(.75,-.5)--++(0,-.1);
\draw[fill=black!100] (\chxbd+1,\chybd+2.4) circle (.07) ;
\draw[fill=black!100] (\chxbd+1.5,\chybd+2.4) circle (.07) ;
\draw[fill=black!100] (\chxbd+2,\chybd+2.4) circle (.07) ;
\node [] at (\chxbd+1.6,\chybd+3.9) {\tiny ${M}_{k,n}$};

\draw[thick,fill=green!10] (\chxcd,\chycd) rectangle ++(3,2) node[midway]{\tiny $U_{k}^{N-1}$};
\draw[thick] (\chxcd+.5,\chycd+2)--+(0,.7);
\draw[thick] (\chxcd+.5,\chycd+2.7)--++(.75,0.5)--++(-1.5,0)--++(.75,-.5)--++(0,-.1);
\draw[thick] (\chxcd+2.5,\chycd+2)--+(0,.7);
\draw[thick] (\chxcd+2.5,\chycd+2.7)--++(.75,0.5)--++(-1.5,0)--++(.75,-.5)--++(0,-.1);
\draw[fill=black!100] (\chxcd+1,\chycd+2.4) circle (.07) ;
\draw[fill=black!100] (\chxcd+1.5,\chycd+2.4) circle (.07) ;
\draw[fill=black!100] (\chxcd+2,\chycd+2.4) circle (.07) ;
\node [] at (\chxcd+1.6,\chycd+3.9) {\tiny ${M}_{k,N-1}$};

\draw[dashed,fill=black!100] (\chxad+3.7,\chyad+2) circle (.1) ;
\draw[dashed,fill=black!100] (\chxad+4,\chyad+1.6) circle (.1) ;
\draw[dashed,fill=black!100] (\chxad+4.3,\chyad+1.2) circle (.1) ;

\draw[dashed,fill=black!100] (\chxbd+.9,\chybd-.9) circle (.1) ;
\draw[dashed,fill=black!100] (\chxbd+.6,\chybd-1.3) circle (.1) ;
\draw[dashed,fill=black!100] (\chxbd+.3,\chybd-1.7) circle (.1) ;

\draw[dashed,black!100] (\chxd+6,\chyd+2) circle (7) ;
\draw[thick,fill=blue!20] (\chxs,\chys) rectangle ++(3,2) node[midway]{\tiny $U_{K}^{N}$};
\draw[thick] (\chxs+.5,\chys+2)--+(0,.7);
\draw[thick] (\chxs+.5,\chys+2.7)--++(.75,0.5)--++(-1.5,0)--++(.75,-.5)--++(0,-.1);
\draw[thick] (\chxs+2.5,\chys+2)--+(0,.7);
\draw[thick] (\chxs+2.5,\chys+2.7)--++(.75,0.5)--++(-1.5,0)--++(.75,-.5)--++(0,-.1);
\draw[fill=black!100] (\chxs+1,\chys+2.4) circle (.07) ;
\draw[fill=black!100] (\chxs+1.5,\chys+2.4) circle (.07) ;
\draw[fill=black!100] (\chxs+2,\chys+2.4) circle (.07) ;
\node [] at (\chxs+1.6,\chys+3.9) {\tiny ${M}_{K,N}$};

\draw[thick,fill=green!10] (\chxas,\chyas) rectangle ++(3,2) node[midway]{\tiny $U_{K}^{1}$};
\draw[thick] (\chxas+.5,\chyas+2)--+(0,.7);
\draw[thick] (\chxas+.5,\chyas+2.7)--++(.75,0.5)--++(-1.5,0)--++(.75,-.5)--++(0,-.1);
\draw[thick] (\chxas+2.5,\chyas+2)--+(0,.7);
\draw[thick] (\chxas+2.5,\chyas+2.7)--++(.75,0.5)--++(-1.5,0)--++(.75,-.5)--++(0,-.1);
\draw[fill=black!100] (\chxas+1,\chyas+2.4) circle (.07) ;
\draw[fill=black!100] (\chxa+1.5,\chyas+2.4) circle (.07) ;
\draw[fill=black!100] (\chxas+2,\chyas+2.4) circle (.07) ;
\node [] at (\chxas+1.6,\chyas+3.9) {\tiny ${M}_{K,1}$};

\draw[thick,fill=green!10] (\chxbs,\chybs) rectangle ++(3,2) node[midway]{\tiny $U_{K}^{n}$};
\draw[thick] (\chxbs+.5,\chybs+2)--+(0,.7);
\draw[thick] (\chxbs+.5,\chybs+2.7)--++(.75,0.5)--++(-1.5,0)--++(.75,-.5)--++(0,-.1);
\draw[thick] (\chxbs+2.5,\chybs+2)--+(0,.7);
\draw[thick] (\chxbs+2.5,\chybs+2.7)--++(.75,0.5)--++(-1.5,0)--++(.75,-.5)--++(0,-.1);
\draw[fill=black!100] (\chxbs+1,\chybs+2.4) circle (.07) ;
\draw[fill=black!100] (\chxbs+1.5,\chybs+2.4) circle (.07) ;
\draw[fill=black!100] (\chxbs+2,\chybs+2.4) circle (.07) ;
\node [] at (\chxbs+1.6,\chybs+3.9) {\tiny ${M}_{K,n}$};

\draw[thick,fill=green!10] (\chxcs,\chycs) rectangle ++(3,2) node[midway]{\tiny $U_{K}^{N-1}$};
\draw[thick] (\chxcs+.5,\chycs+2)--+(0,.7);
\draw[thick] (\chxcs+.5,\chycs+2.7)--++(.75,0.5)--++(-1.5,0)--++(.75,-.5)--++(0,-.1);
\draw[thick] (\chxcs+2.5,\chycs+2)--+(0,.7);
\draw[thick] (\chxcs+2.5,\chycs+2.7)--++(.75,0.5)--++(-1.5,0)--++(.75,-.5)--++(0,-.1);
\draw[fill=black!100] (\chxcs+1,\chycs+2.4) circle (.07) ;
\draw[fill=black!100] (\chxcs+1.5,\chycs+2.4) circle (.07) ;
\draw[fill=black!100] (\chxcs+2,\chycs+2.4) circle (.07) ;
\node [] at (\chxcs+1.6,\chycs+3.9) {\tiny ${M}_{K,N-1}$};

\draw[dashed,fill=black!100] (\chxas+3.7,\chyas+2) circle (.1) ;
\draw[dashed,fill=black!100] (\chxas+4,\chyas+1.6) circle (.1) ;
\draw[dashed,fill=black!100] (\chxas+4.3,\chyas+1.2) circle (.1) ;

\draw[dashed,fill=black!100] (\chxbs+.9,\chybs-.9) circle (.1) ;
\draw[dashed,fill=black!100] (\chxbs+.6,\chybs-1.3) circle (.1) ;
\draw[dashed,fill=black!100] (\chxbs+.3,\chybs-1.7) circle (.1) ;

\draw[dashed,black!100] (\chxs+6,\chys+2) circle (7) ;

\draw[dashed,fill=black!100] (\chxa+8+1,\chya-2-.7) circle (.1) ;
\draw[dashed,fill=black!100] (\chxa+8+1+.5,\chya-2-.7-.35) circle (.1) ;
\draw[dashed,fill=black!100] (\chxa+8+1+1,\chya-2-.7-.7) circle (.1) ;

\draw[dashed,fill=black!100] (\chxbs+4+1,\chybs+2+.7) circle (.1) ;
\draw[dashed,fill=black!100] (\chxbs+4+1+.5,\chybs+2+.7+.35) circle (.1) ;
\draw[dashed,fill=black!100] (\chxbs+4+1+1,\chybs+2+.7+.7) circle (.1) ;
\draw[->,line width=1,green!100] (7.3+.5,-2.2-3)--+(5.5,0);
\draw[->,line width=1,blue!100] (9.3+.5,-3.4-3)--+(3.6,0);
\draw[->,line width=1,red!100] (7.1+.5,-4.6-3-.05)--+(5.7,0);
\node [] at (4.3+.5,-2.2-3) {\scriptsize Energy flow};
\node [] at (5.3+.5,-3.4-3) {\scriptsize Information flow};
\node [] at (4.2+.5,-4.6-3) {\scriptsize Interference};

\draw[->,line width=.6] (31.5,3)--+(1.4,-6.5);
\node [] at (32.8,-4.3) {\scriptsize Cluster Head};
\draw[->,line width=.6] (40.5,3)--+(1,-4.5);
\node [] at (41.5,-2.3) {\scriptsize User};
\end{tikzpicture}

\caption{A ${KN}$-user multi-cluster WPCN with user cooperation.}
\label{ht}
\centering
\end{figure}
We consider a multi-cluster MIMO WPCN consisting of a HAP and ${KN}$ cooperating users denoted by $U_{k}^{n}, 1 \leq k \leq {K}, 1 \leq n \leq {N}$, in ${K}$ clusters as shown in Fig. \ref{ht}. The HAP and user $U_{k}^{n}$ have ${M}_h$ and  ${M}_{k,n}$ antennas, respectively, and all of them operate in the same frequency band. We assume that the HAP has a sustained power supply, whereas the users have no fixed batteries \cite{ju2014user}. Therefore, the HAP shall first broadcast radio frequency (RF) energy to enable users to transmit their information to the HAP by using the energy collected during the downlink phase. We assume that the CH of $k$th cluster denoted by $U_{k}^{N}$ (which can be one of users) is tasked to relay the information of its cluster members $U_{k}^{n}, 1 \leq  n \leq {N-1}$. We further assume that the channels are reciprocal and follow a flat block fading model. We denote the channel response matrix from HAP to the user $U_k^n$ and the channel matrix between $U_i^n$ and the CH, i.e., $U_k^{N}$, respectively by $\mathbf{H}_{k,n}\in \complexC ^{{M}_{k,n} \times {M}_h}, 1 \leq k \leq K, 1 \leq n \leq N$ and $\mathbf{G}_{k,n,i}\in \complexC ^{  {M}_{k,N} \times {M}_{i,n}}, 1 \leq k,i \leq K, 1 \leq n \leq N-1$. We assume that the entries of $\mathbf{H}_{k,n}$ and $\mathbf{G}_{k,n,i}$ are independent random variables with zero mean and variances $\sigma^{2}_{{h}_{k,n}}$ and $\sigma^{2}_{g_{k,n,i}}$, respectively.

\begin{figure}
 	\centering
 	\begin{tikzpicture}
 	\draw[<->,thick] (-7.9,-7.2)--++(1.8,0);
 	\node [] at (-7,-7.5) {$\tau_1$};
 	\draw[<->,thick] (-6.1,-7.2)--++(1.8,0);
 	\node [] at (-5.2,-7.5) {$\tau_2$};
 	\draw[<->,thick] (-4.3,-7.2)--++(1.8,0);
 	\node [] at (-3.4,-7.5) {$\tau_{3,1}$};
 	\draw[<->,thick] (-1.5,-7.2)--++(1.8,0);
 	\node [] at (-.6,-7.5) {$\tau_{3,N-1}$};
 	\draw[<->,thick] (.3,-7.2)--++(2.2,0);
 	\node [] at (1.4,-7.5) {$\tau_{4,1}$};
 	\draw[<->,thick] (3.5,-7.2)--++(2.2,0);
 	\node [] at (4.6,-7.5) {$\tau_{4,N-1}$};
 	\draw[<->,thick] (5.7,-7.2)--++(2.6,0);
 	\node [] at (7,-7.5) {$\tau_{4,N}$};
 	
 	\draw[<->,thick] (-6.1,-4.7)--++(1.8,0)  ;
 	\node [] at (-5.2,-4.3) {\small Downlink};
 	\draw[<->,thick] (-4.3,-4.7)--++(4.6,0)  ;
 	\node [] at (-2,-4.3) {\small Uplink $\tau_3$};
 	\draw[<->,thick] (.3,-4.7)--++(8,0)  ;
 	\node [] at (4.3,-4.3) {\small Uplink $\tau_4$};
 	
 	
 	\draw[thick,fill=brown!15] (-7.9,-7) rectangle ++(1.8,2) node[midway]{\footnotesize \shortstack[l]{ \hspace{2pt} Channel  \\ Estimation \\ \hspace{13pt} \& \\ Clustering}} ;
 	\draw[thick,fill=green!15] (-6.1,-7) rectangle ++(1.8,2) node[midway]{\footnotesize \shortstack[l]{ \hspace{-2.5pt} HAP   \hspace{4.5pt}   ${U}_{k}^{n}$ \\ \hspace{-.15cm}    $k=1,...,{K}$ \\ \hspace{-.14cm}     $n=1,...,{N}$}} ;
 	\draw[->,thick] (-5.29,-5.64)--++(.3,0);	
 	
 	\draw[thick,fill=blue!15] (-4.3,-7) rectangle ++(1.8,2) node[midway]{\footnotesize \shortstack[l]{\hspace{0pt} ${U}_{k}^{1}$   \hspace{5pt}    ${U}_{k}^{N}$ \\      $k=1,...,{K}$}} ;
 	\draw[->,thick] (-3.63,-5.84)--++(.3,0);
 	
 	\draw[thick,fill=blue!15] (-2.5,-7) rectangle ++(1,2) node[midway]{\footnotesize $\hdots$} ;
 	
 	\draw[thick,fill=blue!15] (-1.5,-7) rectangle ++(1.8,2) node[midway]{\footnotesize \shortstack[l]{ ${U}_{k}^{N-1}$   \hspace{1pt}    ${U}_{k}^{N}$ \\      $k=1,...,{K}$}} ;
 	\draw[->,thick] (-.74,-5.84)--++(.32,0);
 	
 	\draw[thick,fill=red!15] (.3,-7) rectangle ++(2.2,2) node[midway]{\footnotesize \shortstack[l]{ \hspace{1pt} ${U}_{k}^{N}$   \hspace{1.5pt}  HAP \\ \hspace{0pt}     $k=1,...,{K}$ \\ (${U}_{k}^{1}$'s signal)}} ;
 	\draw[->,thick] (1.1,-5.62)--++(.32,0);
 	
 	\draw[thick,fill=red!15] (2.5,-7) rectangle ++(1,2) node[midway]{\footnotesize $\hdots$} ;
 	
 	\draw[thick,fill=red!15] (3.5,-7) rectangle ++(2.2,2) node[midway]{\footnotesize \shortstack[l]{ \hspace{5.7pt} ${U}_{k}^{N}$   \hspace{1.5pt}  HAP \\ \hspace{4.5pt}     $k=1,...,{K}$ \\ (${U}_{k}^{N-1}$'s signal)}} ;
 	\draw[->,thick] (1+3.29,-5.62)--++(.32,0);
 	
 	\draw[thick,fill=red!15] (5.7,-7) rectangle ++(2.6,2) node[midway]{\footnotesize \shortstack[l]{ \hspace{11pt} ${U}_{k}^{N}$   \hspace{1.5pt}  HAP \\ \hspace{10pt}     $k=1,...,{K}$ \\ (${U}_{k}^{N}$'s own signal)}} ;
 	\draw[->,thick] (1+3.3+2.38,-5.62)--++(.32,0);

 	\end{tikzpicture}
 	\caption{The proposed multi-cluster cooperation protocol in WPCN.}
 	\label{h1t}
 	\centering
 \end{figure}
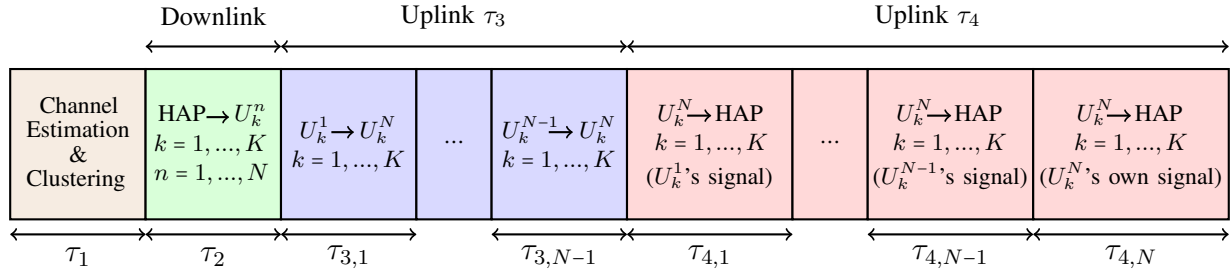

Fig. \ref{h1t} shows the timing protocol of the proposed WPCN which is an enhancement of the HTT protocol in \cite{Throughput}. In this scheme, one complete operation of the system takes $T$ seconds and is divided into four time slots. The first phase/time slot has a fixed duration of $\tau_1$ and is allocated to the channel estimation and user clustering \cite{yuan2017multi}. In this phase, users (except CHs) transmit pilot signals to the HAP and CHs. We model the Linear Minimum Mean Square Error (LMMSE) estimation of $\mathbf{H}_{k,n}$ (at HAP) and $\mathbf{G}_{k,n,i}$ (at CHs) by \cite{kay1993fundamentals1}
 \begin{equation}
 \mathbf{H}_{k,n} = \widehat{\mathbf{H}}_{k,n} + \Delta \mathbf{H}_{k,n}, \hspace{4pt} \forall k,n,
 \label{del1}
 \end{equation}
 \begin{equation}
 {\mathbf{G}}_{k,n,i}= \widehat{{\mathbf{G}}}_{k,n,i} + \Delta {\mathbf{G}}_{k,n.i}
 , \hspace{4pt} \forall k,i,n \neq N,
 \label{del2}
 \end{equation}
 where $\widehat{\mathbf{H}}_{k,n}$ and $\widehat{{\mathbf{G}}}_{k,n,i}$ are the estimates of the channel matrices $\mathbf{H}_{k,n} $ and ${\mathbf{G}}_{k,n,i}$, respectively. Moreover, the channel estimation errors $\Delta \mathbf{H}_{k,n}$ and $\Delta {\mathbf{G}}_{k,n,i}$ are uncorrelated with $\widehat{\mathbf{H}}_{k,n}$ and $\widehat{{\mathbf{G}}}_{k,n,i}$. We also assume that the elements of $\widehat{\mathbf{H}}_{k,n}$, $\widehat{{\mathbf{G}}}_{k,n,i}$, $\Delta \mathbf{H}_{k,n}$ and $\Delta {\mathbf{G}}_{k,n,i}$ are independent random variables with variances $\sigma^{2}_{{\widehat{h}}_{k,n}}$, $\sigma^{2}_{{\widehat{g}}_{k,n,i}}$, $\sigma^{2}_{{h,\Delta}_{k,n}}$ and $\sigma^{2}_{{g,\Delta}_{k,n,i}}$, respectively. The properties of the LMMSE estimator allows to write the following expressions
 \begin{equation}
 \sigma^{2}_{{h,\Delta}_{k,n}} =(1-{\rho}^{2}_{{h}_{k,n}}) \sigma^{2}_{{h}_{k,n}}, \hspace{10pt}\sigma^{2}_{{\widehat{h}_{k,n}}}= {\rho}^{2}_{{h}_{k,n}} \sigma^{2}_{{h}_{k,n}}, \hspace{4pt} \forall k,n,
 \label{sig1}
 \end{equation}
 \begin{equation}
 \sigma^{2}_{{g,\Delta}_{k,n,i}}=(1-{\rho}_{{{g}_{k,n,i}}}^{2}) \sigma^{2}_{{g}_{k,n,i}}, \hspace{10pt}\sigma^{2}_{{\widehat{g}}_{k,n,i}}= {\rho}_{{{g}_{k,n,i}}}^{2} \sigma^{2}_{{g}_{k,n,i}}, \hspace{4pt} \forall k,i,n \neq N,
 \label{sig2}
 \end{equation}
where the parameters $0 \leq {\rho}_{{h}_{k,n}} \leq 1$ and $0 \leq {\rho}_{g_{k,n,i}} \leq 1$ indicate the estimation accuracy. The
 CHs forward the estimates to the HAP and therefore, we assume that the HAP
 has access to these CSI estimates \cite{yuan2017multi}, using which users are clustered by the HAP as a network coordinator (see \cite{na2020clustered,celik2017resource,ding2015impact} for more details).

During the next downlink phase $\tau_{2}$, the HAP emits power to the all users. In the next two uplink phases, the users transmit their information signals to the HAP using their harvested energy. In the third phase, the users of each cluster transmit the information signals to their CH, i.e., $U_k^N$, and the HAP via a time division scheme where $\tau_3 = \sum_{n=1}^{N-1} \tau_{3,n}$. More precisely, in time slot $\tau_{3,n}$, the $n$th users from all clusters simultaneously transmit to the HAP and their CH. During the $\tau_{4,n}$ sub-interval of the fourth phase $\tau_4 = \sum_{n=1}^{N} \tau_{4,n}$, the CHs transmit the decoded signals of $n$th cluster members to the HAP in a similar time division scheme for all $n \neq N$. During $\tau_{4,N}$, the CHs transmit their own signals to the HAP. {All users of each cluster, i.e., $U^{n}_{k}, 1 \leq n \leq N$ could benefit from this cooperation \cite{astaneh2009resource}. This is obvious for $U^{n}_{k},1 \leq n \leq N-1$ but for the CH, i.e., $U^{N}_{k}$, this claim can be verified using the fact that this cooperation allows the HAP to allocate more time for information transmission (i.e., $\tau_3 +\tau_4$) instead of energy transmission (i.e., $\tau_2$) and therefore the CH's throughput loss (due to cooperation) could be compensated by longer uplink time \cite{ju2014user}.}
By dedicating $T$ as the total upper bound operation time of one block, we obtain the following constraint for different phases  \cite{yuan2017multi}
\begin{equation}
\tau_1 + \tau_2 + \sum_{n=1}^{N-1} \tau_{3,n} + \sum_{n=1}^{N} \tau_{4,n}  \leq T.
\label{Qt}
\end{equation}
We now model the signal and system in more details and obtain the link throughputs.
The HAP transfers its energy signal $\mathbf{x} _{0} \in \complexC ^{{M}_h}$  with the EB matrix  $\mathbf{Q} = \mathbb{E} [\mathbf{x}_{0} {\mathbf{x}_{0}}^{{H}}] \succeq \mathbf{0}$ during the downlink phase $\tau_2$ to the users. {We consider $\mathbf{x}_{0}$ as a random vector and design its EB matrix. We discuss the special case of deterministic $\mathbf{x}_{0}$ severalty in Subsection~\ref{det}.}. The transmit power of the HAP is constrained by $ p_{0}$, i.e.,
\begin{equation}
\mathbb{E} [{\| \mathbf{x}_{0} \|}_2^2]  =  \textrm{tr} \lbrace \mathbf{Q} \rbrace  \leq  p_{0}.
\label{e1}
\end{equation}
The received signal at $U_k^{n}$ during $\tau_{2}$ for all $k,n$ is expressed as
\begin{equation}
\mathbf{y}^{(\textrm{HAP-U})}_{k,n} =\left( \widehat{\mathbf{H}}_{k,n} + \Delta {\mathbf{H}}_{k,n} \right) {\mathbf{x}}_{0} + \bz^{(\textrm{HAP-U})}_{k,n},
\end{equation}
where $\bz^{(\textrm{HAP-U})}_{k,n} \sim \mathcal{CN} \left(\mathbf{0},{\mathbf{L}}^{(\textrm{HAP-U})}_{k,n} \right)$ is the received noise with ${\mathbf{L}}^{(\textrm{HAP-U})}_{k,n} \in \mathbb{S}_{++}^{{M}_{k,n}}$. Following \cite{clerckx2018beneficial}, we consider a non-linear circuit model for the EH circuit in which the curve fitting procedure is performed in logarithmic scale to obtain more accurate results in low power regimes \cite{clerckx2019fundamentals}. Therefore, the energy harvested by $U_k^{n}$ (by ignoring the harvested energy due to the receiver noise \cite{yuan2017multi}) for the non-linear EH model (eq. (21) of \cite{clerckx2019fundamentals} in linear scale) for all $k,n$ becomes
\begin{equation}
{E}_{k,n} = \tau_{2} \hspace{1pt} \textrm{exp} \left( a \hspace{1pt} {\left( \textrm{log} \left(   P_{k,n}   \right)  \right)}^2  \right) P^{ \hspace{2pt} b}_{k,n} \hspace{1pt} \textrm{exp} \left( c \right),
\label{fg1nl}
\end{equation}
where $a$, $b$ and $c$ are the curve fitting parameters and
\begin{align} \label{keyp}
P_{k,n} = {\mathbb{E}} \hspace{1pt} \bigg [ {\left( \mathbf{y}^{(\textrm{HAP-U})}_{k,n}\right) }^{{H}} \mathbf{y}^{(\textrm{HAP-U})}_{k,n} \hspace{1pt} \Big \mid \hspace{1pt} \widehat{\mathbf{H}}_{k,n} \bigg]  =
\textrm{tr}  \left \lbrace  {\widehat{\mathbf{H}}_{k,n} }   \mathbf{Q} \hspace{1pt} \widehat{\mathbf{H}}^{{H}}_{k,n}   +   \sigma^{2}_{{h,\Delta}_{k,n}}  \textrm{tr} \left \lbrace \mathbf{Q} \right \rbrace \mathbf{I}_{{M}_{k,n}} \right \rbrace.
\end{align}
During the uplink sub-interval $\tau_{3,n}$, the $n$th users $U_{k}^{n}$ of all clusters transmit their independent message signals to the HAP and their CH by using the energy harvested in the previous phase. Let $\mathbf{V}_{k,n} \in \complexC^{{M}_{k,n} \times {{d}_{k,n}} }$ denote the linear precoder matrix used at $U_{k}^{n}$ to convert the message stream $\mathbf{m}_{k,n} \in \complexC^{{d}_{k,n} }$, viz. a Gaussian random vector with zero mean and  covariance matrix $\mathbf{I}_{{d}_{k,n}}$, to the transmitted vector ${\mathbf{x}}_{k,n} \in \complexC^{{M}_{k,n} }$ with ${\mathbf{x}}_{k,n} =\mathbf{V}_{k,n}  \mathbf{m}_{k,n}$. The total consumed energy at $U_{k}^{n}$ in time slot $\tau_{3,n}$ shall be less than the harvested energy. Thus, we must satisfy the following constraint for all $k$ and $n \neq N$
\begin{equation} \label{5eq}
P_{c_{k,n}}T + \eta_{k,n} \hspace{1pt} {\tau_{3,n}} \hspace{1pt} \textrm{tr} \left \lbrace {\mathbf{{S}}}_{k,n} \right \rbrace \leq  E_{k,n},
 \end{equation}
where $P_{c_{k,n}}$ is the constant circuit power consumption, $\eta_{k,n}$ accounts for power amplifier efficiency and ${\mathbf{{S}}}_{k,n} = \mathbb{E}[ {\mathbf{x}}_{k,n} {\mathbf{x}}^{{H}}_{k,n}] = \mathbf{V}_{k,n} {\mathbf{V}^{{H}}_{k,n}} \in \complexC^{{M}_{k,n} \times {M}_{k,n} }$ is the information transmit covariance matrix at $U_{k}^{n}$.
The received signals at the CH $U_k^N$ and the HAP in time slot $\tau_{3,n}$ for all $k$ and $n \neq N$ are respectively given by
 \begin{align}\label{fg60n}
&\mathbf{y}^{(\textrm{U-CH})}_{k,n} = \widehat{ \mathbf{G}}_{k,n,k}  {\mathbf{x}}_{k,n} + \sum_{j=1,j \neq k}^{{K}}  \widehat{ \mathbf{G}}_{k,n,j}  {\mathbf{x}}_{j,n}  + \sum_{j=1}^{{K}} \Delta {\mathbf{G}}_{k,n,j} {\mathbf{x}}_{j,n} + \mathbf{z}^{(\textrm{U-CH})}_{k,n},
\end {align}
\begin{align}\label{fg60n50}
&\mathbf{y}^{(\textrm{U-HAP})}_n = \sum_{k=1}^{{K}}  \widehat{ \mathbf{H}}^{{H}}_{k,n}  {\mathbf{x}}_{k,n}  + \sum_{k=1}^{{K}} \Delta { \mathbf{H}}^{{H}}_{k,n}  {\mathbf{x}}_{k,n} + {\mathbf{{z}}}^{(\textrm{U-HAP})}_n,
\end{align}
where ${\mathbf{{z}}}^{(\textrm{U-CH})}_{k,n} \sim \mathcal{CN}\left(\mathbf{0},{\mathbf{L}}^{(\textrm{U-CH})}_{k,n} \right)$ and $\mathbf{\widetilde{z}}^{(\textrm{U-HAP})}_n \sim \mathcal{CN}\left(\mathbf{0},{\mathbf{L}}^{(\textrm{U-HAP})}_n \right)$ are the additive noises at the $k$th CH and the HAP, respectively, with covariance matrices ${{\mathbf{L}}}^{(\textrm{U-CH})}_{k,n} \in \mathbb{S}_{++}^{{M}_{k,N}}$ and ${\mathbf{L}}^{(\textrm{U-HAP})}_n \in \mathbb{S}_{++}^{{M}_h} $. The linear decoder matrix $\mathbf{W}_{k,n} \in \complexC^{{d}_{k,n} \times {M}_{k,N}}$ used at $k$th CH estimates the message $\mathbf{m}_{k,n}$  as:
\begin{equation}
\widehat{\mathbf{m}}_{k,n} = \mathbf{W}_{k,n} \mathbf{y}^{(\textrm{U-CH})}_{k,n}, \quad \forall k ~\mbox{ and } \ n \neq N.
\end{equation}
We  express the achievable throughput from $U_{k}^{n}$ to $k$th CH during $\tau_{3,n}$ $\forall k$ and $n \neq N$ as \cite{negro2010mimo}:
\begin{align} \label{newb}
& \hspace{-5pt} R_{k,n}^{(\textrm{U-CH})}\left( \tau_{3,n},{ {\mathbf{{S}}}_{k,n} } ,{\mathbf{W}_{k,n}} \right)  = \tau_{3,n} \hspace{1pt}\textrm{log}_2 \textrm{det} \hspace{1pt} \Bigg(\mathbf{I}_{{d}_{k,n}} +  \mathbf{W}_{k,n} \widehat{{\mathbf{{G}}}}_{k,n,k} {\mathbf{{S}}}_{k,n}  \widehat{{\mathbf{{G}}}}^{{H}}_{k,n,k} \mathbf{W}^{{H}}_{k,n} \bigg(  \mathbf{W}_{k,n} \bigg( {{\mathbf{L}}}^{(\textrm{U-CH})}_{k,n}  \\ \nonumber  & \hspace{130pt} +\sum_{j=1,j \neq k}^{{K}}               \widehat{{\mathbf{{G}}}}_{k,n,j} {\mathbf{{S}}}_{j,n} \widehat{{\mathbf{{G}}}}^{{H}}_{k,n,j} +  \sum_{j=1}^{{K}} \sigma^{2}_{{g,\Delta}_{k,n,j}} \textrm{tr}  \left \lbrace  {\mathbf{{S}}}_{j,n} \right \rbrace \mathbf{I}_{{M}_{k,N}} \bigg)  \mathbf{W}^{{H}}_{k,n} \bigg) ^{-1} \Bigg).
\end{align}
It is shown in \cite{negro2010mimo} the LMMSE is the optimal decoder for interference suppression. Thus, we propose to employ the following LMMSE decoder for all $k$ and $n \neq N$ (see Appendix \ref{app1}):
\begin{equation} \label{w1} {\mathbf{W}}_{k,n}={\mathbf{{V}}}^{{H}}_{k,n}   {\widehat{ \mathbf{G}}^{{H}}_{k,n,k}} \left( {{\mathbf{L}}}^{(\textrm{U-CH})}_{k,n}  + \sum_{j=1}^{{K}} \widehat{ \mathbf{G}}_{k,n,j} {\mathbf{{S}}}_{j,n} \widehat{ \mathbf{G}}^{{H}}_{k,n,j} + \sum_{j=1}^{{K}} \sigma^{2}_{{g,\Delta}_{k,n,j}}  \textrm{tr}  \left \lbrace  {\mathbf{{S}}}_{j,n} \right \rbrace \mathbf{I}_{{M}_{k,N}}  \right) ^{-1}.
\end{equation}
We now use the matrix inversion lemma, Sylvester's determinant property i.e., $\textrm{det} ( \mathbf{I} + \mathbf{A} \mathbf{B}  ) =\textrm{det} (\mathbf{I} + \mathbf{B} \mathbf{A}  )$ and substitute \eqref{w1} in \eqref{newb} in order to rewrite \eqref{newb} as follows for all $k$ and $n \neq N$
\begin{align} \label{fg6}
& \hspace{-5pt}R_{k,n}^{(\textrm{U-CH})}\left( \tau_{3,n}, {\mathbf{{S}}}_{k,n} \right)=
\tau_{3,n}\textrm{log}_2 \textrm{det} \hspace{2pt} \Bigg(\mathbf{I}_{{M}_{k,N}} +  \widehat{ \mathbf{G}}_{k,n,k} {\mathbf{{S}}}_{k,n}  \widehat{ \mathbf{G}}^{{H}}_{k,n,k} \bigg({{\mathbf{L}}}_{k,n}^{(\textrm{U-CH})}  \\ \nonumber & \hspace{100pt}+\sum_{j=1,j \neq k}^{{K}}               \widehat{ \mathbf{G}}_{k,n,j} {\mathbf{{S}}}_{j,n}  \widehat{ \mathbf{G}}^{{H}}_{k,n,j} +  \sum_{j=1}^{{K}} \sigma^{2}_{{g,\Delta}_{k,n,j}} \textrm{tr}  \left \lbrace  {\mathbf{{S}}}_{j,n} \right \rbrace \mathbf{I}_{{M}_{k,N}} \bigg) ^{-1} \Bigg).
\end{align}
The signal transmitted by $U_{k}^{n}$ is also received by the HAP as in \eqref{fg60n50}. For similar arguments, we propose to employ the LMMSE decoder and Successive Interference Cancellation (SIC) technique to decode the messages of the users at the HAP. In this case, we can write the achievable transmission throughput from $U_{k}^{n}$ to the HAP for all $n \neq N$ as \cite{tse2005fundamentals}:
\begin{align} \label{fg10102}
\hspace{-10pt} R_{k,n}^{(\textrm{U-HAP})} \left( \tau_{3,n}, {\mathbf{{S}}}_{k,n}  \right) =
\begin{cases}
\tau_{3,n}\textrm{log}_2 \textrm{det} \hspace{1pt}\Big(\mathbf{I}_{{M}_h} +   \widehat{\mathbf{H}}^{{H}}_{k,n} {\mathbf{{S}}}_{k,n}  \widehat{\mathbf{H}}_{k,n} \big({\mathbf{L}}_{n}^{(\textrm{U-HAP})}  +\sum_{j= k+1}^{{K}}               \widehat{\mathbf{H}}^{{H}}_{j,n} {\mathbf{{S}}}_{j,n}  \widehat{\mathbf{H}}_{j,n}   \\ \hspace{60pt} + \sum_{j=1}^{{K}} \sigma^{2}_{{h,\Delta}_{j,n}} \textrm{tr} \left \lbrace  {\mathbf{{S}}}_{j,n} \right \rbrace \mathbf{I}_{{M}_h} \big) ^{-1} \Big), ~~\forall k \neq K,
\\
\tau_{3,n}\textrm{log}_2 \textrm{det} \hspace{1pt} \Big(\mathbf{I}_{{M}_h} + \widehat{\mathbf{H}}^{{H}}_{k,n} {\mathbf{{S}}}_{k,n}  \widehat{\mathbf{H}}_{k,n} \big({\mathbf{L}}_{n}^{(\textrm{U-HAP})}  \\ \hspace{60pt} +  \sum_{j=1}^{{K}} \sigma^{2}_{{h,\Delta}_{j,n}} \textrm{tr} \left \lbrace  {\mathbf{{S}}}_{j,n} \right \rbrace \mathbf{I}_{{M}_h}\big)^{-1} \Big), ~~ \forall k=K.
\end{cases}
\end{align}
During $\tau_{4,n}$, $U_{k}^{N}$ relays the signal of its $n$th cluster member $U_{k}^{n}$ along with its own signal (for $n=N$) to the HAP.
We express the transmitted vectors of $U_{k}^{N}$ during $\tau_{4,n}$ as follows for all $k,n$:
\begin{equation}
\widetilde{{\mathbf{x}}}_{k,n} =\widetilde{\mathbf{V}}_{k,n}  \widetilde{\mathbf{m}}_{k,n},
\end{equation}
where $\widetilde{{\mathbf{m}}}_{k,n} \in \complexC^{ {{d}_{k,n}}}$ is the symbol stream and $\widetilde{\mathbf{V}}_{k,n} \in \complexC^{{M}_{k,N} \times {{d}_{k,n}} }$ is the precoder employed by $U_{k}^{N}$ for relaying the signals of its $n$th cluster member (for all $n \neq N$) and its own signal (for $n=N$).
We further assume that the elements of messages are independent Gaussian random variables with zero mean and unit variance.
The total energy consumed at the $k$th CH $U_{k}^{N}$ during $\tau_{4}=\sum_{n=1}^{N} \tau_{4,n}$ shall be upper bounded by its harvested energy in the downlink phase. Thus, the following constraint must be satisfied for all $k$
\begin{equation}
P_{c_{k,N}} T + \eta_{k,N} \sum_{n=1}^{N} \tau_{4,n}  \textrm{tr}\lbrace \widetilde{{\mathbf{{S}}}}_{k,n}\rbrace  \leq E_{k,N},
\label{e3}
\end{equation}
where $\widetilde{{\mathbf{{S}}}}_{k,n} = \mathbb{E}[ \widetilde{{\mathbf{x}}}_{k,n} \widetilde{{\mathbf{x}}}^{{H}}_{k,n}] = \widetilde{\mathbf{V}}_{k,n} \widetilde{{\mathbf{V}}}^{{H}}_{k,n} \in \complexC^{{M}_{k,N} \times {M}_{k,N} }$ is the transmit covariance matrix at $U_{k}^{N}$.
Therefore, we can express the received signal at the HAP during $\tau_{4,n}$ for all $n$ as
 \begin{equation}
\mathbf{y}^{(\textrm{CH-HAP})}_n = \sum_{j=1}^{{K}}  \mathbf{H}^{{H}}_{j,N}  \widetilde{{\mathbf{x}}}_{j,n}  + \mathbf{z}^{(\textrm{CH-HAP})}_n,
\label{fg8}
\end{equation}
where $\mathbf{\widetilde{z}}^{(\textrm{CH-HAP})}_n \sim \mathcal{CN}\left(\mathbf{0},{{{\mathbf{L}}}}^{(\textrm{CH-HAP})}_n \right)$ denotes the receiver additive noise with ${{\mathbf{L}}}^{(\textrm{CH-HAP})}_{n} \in \mathbb{S}_{++}^{{M}_{h}}$.
Similar to the third phase, we propose to use LMMSE decoder along with SIC technique at the HAP to decode the corresponding symbol streams, during the last phase.
In this case, we can express the achievable throughput for decoding messages of $U_{k}^{n}$ for $1 \leq n \leq N-1$ and $U_{k}^{N}$ in time slot $\tau_{4,n}$ at HAP as
\begin{align} \label{sic2}
\hspace{-23pt} R_{k,n}^{(\textrm{CH-HAP})} \left( \tau_{4,n}, \widetilde{{\mathbf{{S}}}}_{k,n} \right)=
\begin{cases}
\tau_{4,n}\textrm{log}_2 \textrm{det} \Big(\mathbf{I}_{{M}_h} +   \widehat{\mathbf{H}}^{{H}}_{k,N} \widetilde{{\mathbf{{S}}}}_{k,n}  \widehat{\mathbf{H}}_{k,N} \big(\mathbf{L}_{n}^{(\textrm{CH-HAP})} +\sum_{j= k+1}^{{K}}               \widehat{\mathbf{H}}^{{H}}_{j,N} \widetilde{{\mathbf{{S}}}}_{j,n}  \widehat{\mathbf{H}}_{j,N}  \\ \hspace{60pt}+ \sum_{j=1}^{{K}} \sigma^{2}_{{h,\Delta}_{j,N}} \textrm{tr} \left \lbrace  \widetilde{{\mathbf{{S}}}}_{j,n} \right \rbrace \mathbf{I}_{{M}_h} \big) ^{-1} \Big),~~ \forall k \neq K,
\\
\tau_{4,n}\textrm{log}_2 \textrm{det} \Big(\mathbf{I}_{{M}_h} +   \widehat{\mathbf{H}}^{{H}}_{k,N} \widetilde{{\mathbf{{S}}}}_{k,n}  \widehat{\mathbf{H}}_{k,N} \big(\mathbf{L}_{n}^{(\textrm{CH-HAP})} \\ \hspace{60pt}+ \sum_{j=1}^{{K}} \sigma^{2}_{{h,\Delta}_{j,N}} \textrm{tr} \left \lbrace  \widetilde{{\mathbf{{S}}}}_{j,n} \right \rbrace \mathbf{I}_{{M}_h} \big)^{-1} \Big),~~\forall k= {K}.
\end{cases}
\end{align}
By combining (\ref{fg6}), (\ref{fg10102}) and (\ref{sic2}), the achievable throughput of $U_{k}^{n}$ for all $n \neq N$ becomes \cite{ju2014user}:
\begin{align} \label{fg11}
  R_{k,n}^{(\textrm{U})} =
  \textrm{min}\left( R_{k,n}^{(\textrm{U-HAP})}  +R_{k,n}^{(\textrm{CH-HAP})} , R_{k,n}^{(\textrm{U-CH})} \right).
\end{align}
Moreover, the achievable throughput for $k$th CH is given by (\ref{sic2}), i.e.,
\begin{equation}\label{fg1101}
 R_{k,N}^{(\textrm{U})}=R_{k,N}^{(\textrm{CH-HAP})}.
\end{equation}
\section{The Proposed Fair Throughput Optimization Method }\label{min}
In this section, we cast the max-min fairness throughput problem in which we maximize the throughput of all cooperating users by allocating optimal time durations and optimizing energy and information transmit covariance matrices of the users and HAP in different phases.
We formulate this max-min throughput problem by using \eqref{Qt}, \eqref{e1}, \eqref{5eq}, \eqref{e3}, \eqref{fg11} and \eqref{fg1101} as
\begin{eqnarray}\label{maxmin1}
&&\max_{\hm{\tau}  ,{\mathbf{{Q}}},{\mathbf{{S}}}, \widetilde{\mathbf{{S}}} }   \displaystyle \min_{\substack {{1 \leq k \leq  {K} }, {1 \leq n \leq  {N}}} }  {R}^{(\textrm{U})}_{k,n} \\ \nonumber
\mbox{s.t.}\;\;&& \textrm{C}_{1}:\mathbf{0} \leq \hm{\tau} \leq \mathbf{1}, \hspace{5pt}  \tau_1 + \tau_2 + \sum_{n=1}^{N-1}\tau_{3,n} + \sum_{n=1}^{N}\tau_{4,n}  \leq 1, \hspace{80pt}  \\ \nonumber
&& \textrm{C}_{2}: {\mathbf{{Q}}} \succeq \mathbf{0} , \hspace{3pt} \textrm{tr} \left \lbrace {\mathbf{{Q}}} \right \rbrace \leq p_{0},\\ \nonumber \;\;&&
\textrm{C}_{3}: {\mathbf{{S}}}_{k,n} \succeq \mathbf{0}, \hspace{3pt} P_{c_{k,n}} +  \eta_{k,n}  {\tau_{3,n}}\textrm{tr} \lbrace {\mathbf{{S}}}_{k,n} \rbrace \leq  E_{k,n} , \hspace{3pt}  \forall k, n \neq N, \\ \nonumber \;\;&&
\textrm{C}_{4}: \widetilde{{\mathbf{{S}}}}_{k,n} \succeq \mathbf{0}, \hspace{3pt} P_{c_{k,N}} + \eta_{k,N} \sum_{n=1}^{N}\tau_{4,n}   \textrm{tr} \lbrace \widetilde{{\mathbf{{S}}}}_{k,n} \rbrace \leq  E_{k,N} , \hspace{3pt}  \forall k,
\end{eqnarray}
where $\mathbf{S} = [\mathbf{S}_{k,n},\hspace{2pt}\forall k, n \neq N]$, $\widetilde{\mathbf{S}} = [\widetilde{\mathbf{S}}_{k,n},\hspace{2pt}\forall k,n]$, $\hm{\tau}=[\tau_{2},\tau_{3,1}, ..., \tau_{3,N-1},\tau_{4,1}, ..., \tau_{4,N}]^{{T}}$ and without loss of generality, we normalize the total time to $T = 1$. We note that the problem in (\ref{maxmin1}) is not convex due to the coupled design variables in the objective function and in the constraints $\textrm{C}_{3}$ and $\textrm{C}_{4}$. To deal with this problem, we devise a method based on AO with partitioning $[\hm{\tau} \hspace{5pt} \vdots \hspace{5pt} \mathbf{{Q}}, \mathbf{{S}}, \widetilde{\mathbf{{S}}}]$. We first solve the problem for $[\mathbf{{Q}}, \mathbf{{S}}, \widetilde{\mathbf{{S}}}]$ assuming a fixed solution for $\hm{\tau}$. Then, we solve the problem for $\hm{\tau}$ assuming a fixed solution for $[\mathbf{{Q}}, \mathbf{{S}}, \widetilde{\mathbf{{S}}}]$. These two steps will continue till a predefined stop criterion is satisfied for the convergence.
\subsection{Maximization w.r.t. Transmit Covariance Matrices} \label{keymat}
We recast the problem in (\ref{maxmin1}) by introducing an auxiliary variable ${\beta_{a}}$ and using \eqref{fg11}-\eqref{fg1101}  as follows
\begin{eqnarray}\label{maxmin101}
\max_{\mathbf{Q},\mathbf{S},\widetilde{\mathbf{S}}, \beta_{a} } & & {\beta_{a}}  \\ \nonumber
\mbox{s.t.}\;\;&& \textrm{C}^{a}_{5,1}: R_{k,n}^{(\textrm{U-CH})}(\mathbf{S})   \geq {\beta_{a}},  \hspace{10pt} \textrm{C}^{a}_{5,2}: R_{k,n}^{(\textrm{U-HAP})}(\mathbf{S}) +R_{k,n}^{(\textrm{CH-HAP})}(\widetilde{\mathbf{S}})  \geq {\beta_{a}}, \hspace{3pt}  \forall k,n \neq N, \\ \nonumber \;\;&&
\textrm{C}^{a}_{6}: {R}^{(\textrm{CH-HAP})}_{k,N}(\widetilde{\mathbf{S}}) \geq {\beta_{a}}, \hspace{3pt}  \forall k, \hspace{10pt}
\textrm{C}_{2} - \textrm{C}_{4}.
\end{eqnarray}
We observe that the constraints $\textrm{C}^{a}_{5,1}$, $\textrm{C}^{a}_{5,2}$ and $\textrm{C}^{a}_{6}$ in \eqref{maxmin101} are not convex and so the problem in \eqref{maxmin101}. Thus, in what follows, we use the MM technique to devise an algorithm to solve this non-convex problem.

The iterative MM can be used to obtain a solution for some general optimization problems:
\begin{align}\label{mami1}
P_0: \hspace{2pt} \left\{
\begin{array}{ll}
&\underset{{\bx}} {\max}\;\; \widetilde{f}({\bx}) \\
& \textrm{s.t.} \quad \widetilde{g}({\bx})\leq 0,
\end{array}
\right.
\end{align}
where $\widetilde{f}({\bx})$ and $\widetilde{g}({\bx})$ may be non-convex functions. To apply MM in $P_0$, we shall find two functions $\widetilde{h}^{(\kappa)}({\bx})$ and $\widetilde{q}^{(\kappa)}({\bx})$ at $\kappa$th iteration such that $\widetilde{q}^{(\kappa)}({\bx})$ minorizes $\widetilde{f}({\bx})$, i.e.:
\begin{align}\label{mami3}
   &\widetilde{f}({\bx}) \geq \widetilde{q}^{(\kappa)}({\bx}) ,\hspace{1.5pt} \forall {\bx}, \hspace{15pt} \widetilde{f}({\bx}^{(\kappa-1)})=\widetilde{q}^{(\kappa)}({\bx}^{(\kappa-1)}),
\end{align}
and $\widetilde{h}^{(\kappa)}({\bx})$ majorizes $\widetilde{g}({\bx})$ as for all ${\bx}$ follows
\begin{align}\label{mami2}
   &\widetilde{h}^{(\kappa)}({\bx})\geq \widetilde{g}({\bx}), \hspace{15pt} \widetilde{h}^{(\kappa)}({\bx}^{(\kappa-1)})=\widetilde{g}({\bx}^{(\kappa-1)}),
\end{align}
where ${\bx}^{(\kappa-1)}$ is the value of ${\bx}$ at the $(\kappa-1)$th iteration. The following optimization problem is solved at the $\kappa$th iteration (which is simpler than the original problem):
\begin{align}\label{mami33}
P_{\kappa}: \hspace{2pt} \left\{
\begin{array}{ll}
  &\underset{{\bx}} {\max}\;\; \widetilde{q}^{(\kappa)}({\bx}) \\
  & \textrm{s.t.} \quad \widetilde{h}^{(\kappa)}({\bx})\leq 0.
      \end{array}
  \right.
\end{align}
We first apply MM technique on $\textrm{C}^b_{5,1}$ by using the matrix inversion lemma and rewriting $R_{k,n}^{(\textrm{U-CH})}$ in \eqref{fg6} for all $k$ and $n \neq N$ (see Appendix~\ref{appnew}) as
\begin{equation}
 R_{k,n}^{(\textrm{U-CH})}=  \tau_{3,n}\textrm{log}_2 \textrm{det} \left(\mathbf{A}^ {{H}}_{k,n}  \mathbf{D}_{k,n}^{-1}\mathbf{A}_{k,n} \right),
\label{fg60}
\end{equation}
where
 \begin{equation}
 \mathbf{A}_{k,n}=[\mathbf{I}_{{M}_{k,n}}, \mathbf{0}_{{M}_{k,n} \times {M}_{k,N}}]^{T},
\label{fg61}
\end{equation}
\begin{equation} \label{2111}
\mathbf{{D}}_{k,n} =
\begin{bmatrix}
\mathbf{I}_{{M}_{k,n}}  \hspace*{.5mm}&\hspace*{.5mm}   ({\mathbf{{S}}}^{\frac{1}{2}}_{k,n} )^{{H}} {{\widehat{\mathbf{{G}}}}^{{H}}_{k,n,k}}   \vspace*{3mm} \\
{\widehat{\mathbf{{G}}}}_{k,n,k}  {{\mathbf{{S}}}^{\frac{1}{2}}_{k,n}} \hspace*{.5mm}&\hspace*{.5mm}  {{\mathbf{L}}}_{k,n}^{(\textrm{U-CH})} +\sum_{j=1}^{{K}} \left \lbrace  {{\widehat{\mathbf{{G}}}}_{k,n,j}}  {\mathbf{S}}_{j,n}   {\widehat{\mathbf{{G}}}}^{{H}}_{k,n,j} +  \sigma^{2}_{{g,\Delta}_{k,n,j}} \textrm{tr}  \left \lbrace  {\mathbf{{S}}}_{j,n} \right \rbrace \mathbf{I}_{{M}_{k,N}} \right \rbrace
\end{bmatrix}.
\end{equation}
Lemma~1 from \cite{naghsh2017information} allows us to find a minorizer for $R_{k,n}^{(\textrm{U-CH})}$ in \eqref{fg60}.

$\textbf{Lemma 1.}$ The function $f( \mathbf{X})= \textrm{log}_2 \textrm{det} \left(\mathbf{A}^ H  {\mathbf{X}}^{-1}\mathbf{A} \right)$ is convex in $\mathbf{X} \in \mathbb{S}_{++}^{N}$ for any full column rank matrix $\mathbf{A}$.
$\hspace{388pt} \blacksquare$

It can be proved that $ \mathbf{{D}}_{k,n}$ for all $k$ and $n \neq N$ is positive definite by using schur complement \cite{meyer2000matrix}. Thus, Lemma 1 implies that $R_{k,n}^{(\textrm{U-CH})}$ is convex w.r.t. $ \mathbf{{D}}_{k,n}$. This gives the following minorization at $\kappa$th iteration of the algorithm \cite{ naghsh2017information,dattorro2010convex } for all $k$ and $n \neq N$:
\begin{equation}
 \tau_{3,n}\textrm{log}_2 \textrm{det} \left(\mathbf{A}_{k,n} ^{{H}}   \mathbf{D}_{k,n}^{-1}\mathbf{A}_{k,n} \right) \geq
 \tau_{3,n}\textrm{log}_2 \textrm{det} \left(\mathbf{A}_{k,n}^{{H}}  \left(\mathbf{D}^{(\kappa-1)}_{k,n} \right)^{-1}\mathbf{A}_{k,n} \right)  - \tau_{3,n} \textrm{tr} \left \lbrace \mathbf{F}^{(\kappa-1)}_{k,n} \left(\mathbf{D}_{k,n} - \mathbf{D}^{(\kappa-1)}_{k,n} \right) \right \rbrace,
\label{fg600}
\end{equation}
where $\mathbf{F}^{(\kappa-1)}_{k,n}= \left(\mathbf{D}^{(\kappa-1)}_{k,n} \right)^{-1}\mathbf{A}_{k,n} \left(\mathbf{A}^{{H}}_{k,n}  \left(\mathbf{D}^{(\kappa-1)}_{k,n} \right)^{-1} \mathbf{A}_{k,n} \right)^{-1} \mathbf{A}^{{H}}_{k,n}  \left(\mathbf{D}^{(\kappa-1)}_{k,n} \right)^{-1}  \succeq \mathbf{0}$.

We now derive an explicit expression for the constraint $\textrm{C}^{a}_{5,1}$ in terms of the design variables $\mathbf{S}_{k,n}$ by partitioning
\begin{equation} \label{2041}
\mathbf{F}_{k,n} =
\begin{bmatrix}
(\mathbf{F}_{k,n})_{11} \in \complexC^ {{{M}_{k,n} \times {M}_{k,n}}}  \hspace*{3mm}&\hspace*{3mm}   (\mathbf{F}_{k,n})_{12} \in \complexC^ {{{M}_{k,n} \times {M}_{k,N}}}  \vspace*{3mm} \\
(\mathbf{F}_{k,n})_{21} \in \complexC^ {{{M}_{k,N} \times {M}_{k,n}}}  \hspace*{3mm}&\hspace*{3mm}  (\mathbf{F}_{k,n})_{22} \in \complexC^ {{{M}}_{k,N} \times {M}_{k,N}}
\end{bmatrix}.
\end{equation}
Next by using \eqref{2111}, we can write the following expression for all $k$ and $n \neq N$:
\begin{align}
 \textrm{tr} \left \lbrace \mathbf{F}_{k,n} \mathbf{D}_{k,n} \right \rbrace & = 2 \Re \left\lbrace \textrm{tr} \left \lbrace {{(\mathbf{F}_{k,n})}_{12}} {\widehat{\mathbf{{G}}}}_{k,n,k} {\mathbf{{S}}}^{\frac{1}{2}}_{k,n} \right\rbrace  \right\rbrace  +\textrm{tr} \left\lbrace {(\mathbf{F}_{k,n})}_{11} \right\rbrace+\textrm{tr} \left\lbrace {(\mathbf{F}_{k,n})_{22}} {{\mathbf{L}}}^{\textrm{(U-CH)}}_{k,n}  \right\rbrace  \\ \nonumber & \hspace{5pt}
  +\textrm{tr} \left\lbrace { (\mathbf{F}_{k,n})}_{22} \sum_{j=1}^{{K}} \left \lbrace {\widehat{\mathbf{{G}}}}_{k,n,j} {{\mathbf{{S}}}_{j,n}}   {{\widehat{\mathbf{{G}}}}^{{H}}_{k,n,j}}  +  \sigma^{2}_{{g,\Delta}_{k,n,j}} \textrm{tr}  \left \lbrace  {\mathbf{{S}}}_{j,n} \right \rbrace \mathbf{I}_{{M}_{k,N}} \right\rbrace \right\rbrace.
\label{fg6000}
\end{align}
We rewrite the right-hand side of \eqref{fg600} by defining $\mathbf{s}_{k,n} \triangleq \textrm{vec}({{\mathbf{{S}}}^{\frac{1}{2}}_{k,n}})$ and using $\textrm{tr}   \lbrace \mathbf{A} \mathbf{B} \rbrace= (  \textrm {vec} ( \mathbf{A}^{{T}} ))^{{T}}  \textrm {vec}( \mathbf{B} ) $ and $\textrm {vec} ( \mathbf{A} \mathbf{B} \mathbf{C} )= ( \mathbf{C}^{{T}} \otimes \mathbf{A}) \textrm {vec} ( \mathbf{B} )$ for all $k$ and $n \neq N$ as:
\begin{equation}
-\tau_{3,n} \left(\mathbf{T}^{(\kappa-1)}_{k,n} +2 \Re \left \lbrace \left(\mathbf{v}^{(\kappa-1)}_{k,n}\right)^{{H}} \mathbf{s}_{k,n} \right \rbrace +\sum_{j=1}^{{K}} {\mathbf{s}}^{{H}}_{j,n} \mathbf{\Upsilon}^{(\kappa-1)}_{k,n,j} {\mathbf{s}}_{j,n} \right),
\label{fg6001}
\end{equation}
where by considering $\textrm{tr} \left\lbrace { (\mathbf{F}_{k,n})}_{22}  \sigma^{2}_{{g,\Delta}_{k,n,j}} \textrm{tr}  \left \lbrace  {\mathbf{{S}}}_{j,n} \right \rbrace \mathbf{I}_{{M}_{k,N}} \right\rbrace = \textrm{tr} \left\lbrace \textrm{tr} \left\lbrace { (\mathbf{F}_{k,n})}_{22} \right\rbrace  \sigma^{2}_{{g,\Delta}_{k,n,j}}   {\mathbf{{S}}}_{j,n} \right \rbrace, $ we let
\begin{align} \label{fg6a11}
\mathbf{T}^{(\kappa-1)}_{k,n} &= - \textrm{log}_2 \textrm{det} \left(\mathbf{A}_{k,n}^{{H}}\left(\mathbf{D}^{(\kappa-1)}_{k,n} \right)^{-1}\mathbf{A}_{k,n} \right) - \textrm{tr} \left\lbrace\mathbf{F}^{(\kappa-1)}_{k,n} \mathbf{D}^{(\kappa-1)}_{k,n}\right\rbrace \\ \nonumber & \hspace{5pt} +\textrm{tr} \left\lbrace \left(\mathbf{F}^{(\kappa-1)}_{k,n} \right)_{11}\right \rbrace + \textrm{tr} \left \lbrace \left(\mathbf{F}^{(\kappa-1)}_{k,n} \right)_{22} {{\mathbf{L}}}_{k,n}^{\textrm{(U-CH)}} \right\rbrace,
\end{align}
\begin{equation}
 \mathbf{v}^{(\kappa-1)}_{k,n}=\textrm{vec}\left( {{\widehat{\mathbf{{G}}}}^{{H}}_{k,n,k}} \left(\mathbf{F}^{(\kappa-1)}_{k,n} \right)^{{H}}_{12} \right),
\label{fg6b}
\end{equation}
\begin{equation} \mathbf{\Upsilon}^{(\kappa-1)}_{k,n,j}=\mathbf{I}_{{M}_{j,n}}  \otimes \left( {{\widehat{\mathbf{{G}}}}^{{H}}_{k,n,j}} \left(\mathbf{F}^{(\kappa-1)}_{k,n} \right)_{22} {\widehat{\mathbf{{G}}}}_{k,n,j} + \sigma^{2}_{{g,\Delta}_{k,n,j}} \textrm{tr} \left\lbrace { (\mathbf{F}_{k,n})}_{22} \right\rbrace  \mathbf{I}_{{M}_{j,n}} \right).
\label{fg6c}
\end{equation}
We have ${\left(\mathbf{F}^{(\kappa-1)}_{k,n} \right)}_{22}\succeq \mathbf{0}$ because $\mathbf{F}^{(\kappa-1)}_{k,n} \succeq \mathbf{0}$. Thus, ${{\widehat{\mathbf{{G}}}}^{{H}}_{k,n,j}} (\mathbf{F}^{(\kappa-1)}_{k,n} )_{22} {\widehat{\mathbf{{G}}}}_{k,n,j}\succeq \mathbf{0}$. Consequently, from the Kronecker product properties, we have $\mathbf{\Upsilon}^{(\kappa-1)}_{k,n,j}\succeq \mathbf{0}$ since ${{\widehat{\mathbf{{G}}}}^{{H}}_{k,n,j}} \left(\mathbf{F}^{(\kappa-1)}_{k,n} \right)_{22} {\widehat{\mathbf{{G}}}}_{k,n,j} + \sigma^{2}_{{g,\Delta}_{k,n,j}} \textrm{tr} \left\lbrace { (\mathbf{F}_{k,n})}_{22} \right\rbrace  \mathbf{I}_{{M}_{j,n}}\succeq \mathbf{0}$.
Therefore, \eqref{fg6001} is a quadratic and concave expression in terms of ${\mathbf{s}}_{k,n}$.

We manage the constraints $\textrm{C}^{a}_{5,2}$ and $\textrm{C}^{a}_{6}$ in \eqref{maxmin101} similar to $\textrm{C}^{a}_{5,1}$ by applying the MM technique. Precisely, we consider SIC (see \eqref{fg10102} and substitute \eqref{sic2}), the left-hand side of the constraints $\textrm{C}^{a}_{5,2}$ and $\textrm{C}^{a}_{6}$ respectively with the following expressions at the $\kappa$th iteration
\begin{align} \label{fg60101}
& - \tau_{3,n} \left(\widetilde{\mathbf{T}}^{(\kappa-1)}_{k,n} +2 \Re \left\lbrace \left(\widetilde{\mathbf{v}}^{(\kappa-1)}_{k,n} \right)^{{H}} {\mathbf{s}}_{k,n}  \right\rbrace +\sum_{j=k}^{{K}} {\mathbf{s}}^{{H}}_{j,n} \widetilde{\mathbf{\Upsilon}}^{(\kappa-1)}_{j,n} {\mathbf{s}}_{j,n}  \right)
\\ \nonumber &
 - \tau_{4,n} \left(\bar{\mathbf{T}}^{(\kappa-1)}_{k,n} +2 \Re \left\lbrace \left(\bar{\mathbf{v}}^{(\kappa-1)}_{k,n} \right)^{{H}} {\widetilde{\mathbf{s}}}_{k,n} \right\rbrace +\sum_{j=k}^{{K}} {{\widetilde{\mathbf{s}}}^{{H}}_{j,n}} \bar{\mathbf{\Upsilon}}^{(\kappa-1)}_{j,n} {\widetilde{\mathbf{s}}}_{j,n}\right) ,
\end{align}
\begin{equation}
- \tau_{4,N} \left(\bar{\mathbf{T}}^{(\kappa-1)}_{k,N} +2  \Re \left\lbrace \left(\bar{\mathbf{{v}}}^{(\kappa-1)}_{k,N} \right)^{{H}} { \widetilde{\mathbf{s}}}_{k,N} \right\rbrace +\sum_{j=k}^{{K}}  {{\widetilde{\mathbf{s}}}^{{H}}_{j,N}} \bar{\mathbf{\Upsilon}}^{(\kappa-1)}_{j,N} {\widetilde{\mathbf{s}}}_{j,N}\right)  ,
\label{fg60111}
\end{equation}
where ${{\widetilde{\mathbf{{s}}}}_{k,n}} \triangleq \textrm{vec}({{\widetilde{\mathbf{{S}}}}^{\frac{1}{2}}_{k,n}})$. Moreover, the matrices/vectors $\widetilde{\mathbf{T}}^{(\kappa-1)}_{k,n} $,  $\bar{\mathbf{T}}^{(\kappa-1)}_{k,n}$, $\widetilde{\mathbf{v}}^{(\kappa-1)}_{k,n}$,  $\bar{\mathbf{v}}^{(\kappa-1)}_{k,n}$, $\widetilde{\mathbf{\Upsilon}}^{(\kappa-1)}_{j,n}$,  $\bar{\mathbf{\Upsilon}}^{(\kappa-1)}_{j,n}$, $\widetilde{\mathbf{F}}^{(\kappa-1)}_{k,n}$,  $\bar{\mathbf{F}}^{(\kappa-1)}_{k,n}$, $\widetilde{\mathbf{A}}_{k,n} $, $\bar{\mathbf{A}}_{k,n}$, $\widetilde{\mathbf{{D}}}_{k,n}$ and $\bar{\mathbf{{D}}}_{k,n}$ are defined in Appendix \ref{app3}.

Finally, the $\kappa$th iteration of the proposed method is handled via solving the following problem
\begin{eqnarray}\label{maxmin1001}
\max_{{\mathbf{{Q}}}, {\mathbf{{s}}}, \widetilde{\mathbf{s}}, {\beta_{a}}} & & {\beta_{a}}  \\ \nonumber
\mbox{s.t.}\;\;&& \textrm{C}^{a}_{5,1}: \eqref{fg6001} \geq \beta_{a} , \hspace{10pt}  \textrm{C}^{a}_{5,2}: \eqref{fg60101} \geq \beta_{a}, \hspace{10pt}
\textrm{C}^{a}_{6}: \eqref{fg60111}\geq \beta_{a},  \hspace{10pt}
\textrm{C}_{2},  \\ \nonumber \;\;&&
   \textrm{C}^{a}_{3}:
 P_{c_{k,n}} +\eta_{k,n} {\tau_{3,n}}   {\| {\mathbf{{s}}}_{k,n}\|}^2_2 \leq  E_{k,n}  , \hspace{4pt}  \forall k, n \neq N,
  \\ \nonumber \;\;&&   \textrm{C}^{a}_{4}:
  P_{c_{k,N}} +\eta_{k,N} \sum_{n=1}^{N}\tau_{4,n}  {\| {\widetilde{\mathbf{{s}}}}_{k,n}\|}^2_2  \leq E_{k,N}, \hspace{4pt} \forall k,
\end{eqnarray}
where ${\mathbf{s}} =[{\mathbf{s}}_{k,n},\hspace{2pt} \forall k, 1 \leq n \leq N-1]$ and $\widetilde{\mathbf{s}} =[\widetilde{\mathbf{s}}_{k,n},\hspace{2pt} \forall k, n]$. We observed that the problem in \eqref{maxmin1001} is convex and particularly is a SOCP w.r.t. $[{\mathbf{s}}, \widetilde{\mathbf{s}}]$ with linear constraints on matrix $\mathbf{{Q}}$ and $\beta_{a}$. Therefore, \eqref{maxmin1001} can be quickly solved e.g. via interior point methods.
\subsection{Maximization w.r.t. $\hm{\tau}$} \label{keyLP}
We introduce an auxiliary variable $\beta_{b}$ and use \eqref{fg11} and \eqref{fg1101} in (\ref{maxmin1}) to simplify the problem into its epigraphic form as follows
\begin{eqnarray}\label{maxmin10551}
	\max_{\hm{\tau}, {\beta_{b}}} & & {\beta_{b}}  \\ \nonumber
	\mbox{s.t.}\;\;&& \textrm{C}^{b}_{5,1}: R_{k,n}^{(\textrm{U-CH})}(\hm{\tau})   \geq {\beta_{b}},  \hspace{10pt} \textrm{C}^{b}_{5,2}: R_{k,n}^{(\textrm{U-HAP})}(\hm{\tau}) +R_{k,n}^{(\textrm{CH-HAP})}(\hm{\tau})  \geq {\beta_{b}}, \hspace{3pt}  \forall k,n \neq N, \\ \nonumber \;\;&&
	\textrm{C}^{b}_{6}: {R}^{(\textrm{CH-HAP})}_{k,N}(\hm{\tau}) \geq {\beta_{b}}, \hspace{3pt}  \forall k, \hspace{10pt}
	\textrm{C}_{1}, \textrm{C}_{3}, \textrm{C}_{4}.
\end{eqnarray}
It is observed that \eqref{maxmin10551} is a Linear Programming (LP) and can be solved efficiently e.g. by  interior point methods.

Table \ref{table:method2} summaries the steps of this  algorithm. This method has a nested loop. In the outer loop denoted by superscript $l$, we alternate between partitioned variables $[\hm{\tau} \hspace{5pt} \vdots \hspace{5pt} \mathbf{{Q}}, \mathbf{{S}}, {\widetilde{\mathbf{{S}}}}]$. In step 1, $[\mathbf{{Q}},\mathbf{{S}}, {\widetilde{\mathbf{{S}}}}]$ are optimized for fixed $\hm {\tau}$. In step 2, $\hm {\tau}$ is optimized for fixed $[\mathbf{{Q}},\mathbf{{S}}, {\widetilde{\mathbf{{S}}}}]$. Note that optimizing $[\mathbf{{Q}}, \mathbf{{S}}, {\widetilde{\mathbf{{S}}}}]$ for fixed $\hm {\tau}$ is performed with inner iterations denoted by superscript $\kappa$ introduced by the MM to deal with non-convex constraint set in \eqref{maxmin101}. We note that the LP in \eqref{maxmin10551} and SOCP in \eqref{maxmin1001} are solved efficiently by using interior point methods in polynomial time.
\begin{rema}[Convergence]
The sequence of objective values of Problem \eqref{maxmin101} obtained by applying MM in step 1 has a ascent property and converges to a finite value \cite{rezaei2019throughput}. Moreover, the sequence of the objective values of \eqref{maxmin1}, the design problem, has the same property and converges to a finite value (see \cite{rezaei2019throughput} for the conditions in which MM and AO converge to stationary points).
\end{rema}
\begin{rema}[Complexity Analysis]
	The problem in \eqref{maxmin1001} (see step~1 of Algorithm~1) has $2+2K(N-1)$ variables. Also, $\textrm{C}_2$ consists of two sets of constraints, each of dimension $M^2_h$. Moreover, $\textrm{C}^{a}_3$ is associated with $K(N-1)$ constraints with dimension\footnote{This value is computed for the worst case in terms of constraint dimension.} $\displaystyle \max_{k,n\neq N} M^2_{k,n}$. Note that dimensions of $\textrm{C}^{a}_4$, $\textrm{C}^{a}_{5,1}$, $\textrm{C}^{a}_{5,2}$, and $\textrm{C}^{a}_6$ in \eqref{maxmin1001} can be obtained similarly. Therefore, applying interior point methods to the problem in \eqref{maxmin1001} leads to computational complexity of (see \cite{ben2001lectures} for details)
	\begin{equation}
		\displaystyle \mathcal{O}\bigg( \Big( (2+2K(N-1)) (2M^2_h + 3K(N-1) \max_{k,n\neq N} M^2_{k,n} + K(N+1) \max_{k} M^2_{k,N}) \Big)^{3.5} \bigg).
	\end{equation}	
	As to step~2 of Algorithm~1, we note that it is nothing but an LP with computational complexity of $\mathcal{O}\left( \sqrt{2N+1} \right)$ \cite{gondzio1994computational}, where $2N+1$ denotes the number of variables. Note that the total computational complexity of step~1 and 2 are linear with the number of iterations.
\end{rema}
\begin{table}[tp]
\footnotesize
\caption{The Proposed Fair Throughput Maximization Method for a Cooperative Multi-Cluster WPCN} \label{table:method2} \centering
\begin{tabular}{p{4in}}
\hline \hline
\textbf{Step 0}: Initialize $\mathbf{Q}$, ${\mathbf{s}}$ and ${\widetilde{\mathbf{s}}}$ such that satisfy the constraints $\textrm{C}_{2}, \textrm{C}^{b}_{3}$ and $\textrm{C}^{b}_{4}$. \\
\textbf{Step 1}: Compute $\mathbf{Q}^{(l)}$, ${\mathbf{s}}^{(l)}$ and ${\widetilde{\mathbf{s}}}^{(l)}$.\\
~~~\textbf{Step 1-1}: Solve the convex problem in \eqref{maxmin1001}.\\
~~~\textbf{Step 1-2}: Update ${\bT}_{k,n}$, $\widetilde{\bT}_{k,n}$, $\bar{\bT}_{k,n}$, ${\bv}_{k,n}$, $\widetilde{\bv}_{k,n}$, $\bar{\bv}_{k,n}$,${\mathbf{\Upsilon}}_{k,n,j}$, $\widetilde{\mathbf{\Upsilon}}_{j,n}$ and\\~~~$\bar{\mathbf{\Upsilon}}_{j,n}$ according to \eqref{fg6a11}, \eqref{fg6b}, \eqref{fg6c} and Appendix~\ref{app3}.\\
~~~\textbf{Step 1-3}: Repeat steps 1-1 and 1-2 till the stop criterion is satisfied.\\
\textbf{Step 2}: Compute $\hm{\tau}^{(l)}$ by solving the LP in (\ref{maxmin10551}).\\
\textbf{Step 3}: Repeat steps 1 and 2 until a pre-defined stop criterion is satisfied, e.g. $|g^{(l+1)}-g^{(l)}| \leq \xi$ (where $g$ denotes the objective function of the Problem \eqref{maxmin1}) for some $\xi>0$.\\
\hline \hline
\end{tabular}
\end{table}
\section{Extensions} \label{further}
\subsection{Deterministic Energy Signal} \label{det}
Here we consider the energy signal $\mathbf{x}_{0}$ as a deterministic signal with a rank-1 EB matrix $\mathbf{Q} =\mathbf{x} _{0} \mathbf{x}^{{H}} _{0}$. In this case, the instantaneous received energy by $U_k^{n}$ during the downlink phase can be used more reliably for information transmission in the next phases \cite{rezaei2019throughput}.

From \eqref{keyp}, we can express the RF input power of user $U_k^{n}$ as
\begin{align}
	\widetilde{P}_{k,n} (\mathbf{x}_{0})= {\mathbb{E}} \hspace{1pt} \bigg [ {\left( \mathbf{y}^{(\textrm{HAP-U})}_{k,n}\right) }^{{H}} \mathbf{y}^{(\textrm{HAP-U})}_{k,n} \hspace{1pt} \Big \mid \hspace{1pt} \widehat{\mathbf{H}}_{k,n} \bigg]  = \mathbf{x}^{{H}}_{0} \left(\mathbf{B}_{k,n} \right) \mathbf{x}_{0} , \hspace{4pt} \forall k,n,
\end{align}
where $\mathbf{B}_{k,n}=
\widehat{\mathbf{H}}^{{H}}_{k,n}  {\widehat{\mathbf{H}}_{k,n} }    +   \sigma^{2}_{{h,\Delta}_{k,n}} {{M}_{k,n}}  \mathbf{I}_{{M}_{h}}$. The constraints $\textrm{C}_{2}$ and $\textrm{C}^{a}_{3}-\textrm{C}^{a}_{4}$ in Problems \eqref{maxmin1001} and \eqref{sum3} can be respectively updated as
\begin{align*}
	&{\textrm{C}}^{\textrm{det}}_{2}: \| {\mathbf{x}}_{0} \|_2^2 \leq p_{0} ,
\end{align*}
\begin{align*}
	&{\textrm{C}}^{\textrm{det}}_{3}: P_{c_{k,n}} +\eta_{k,n} {\tau_{3,n}}   {\| {\mathbf{{s}}}_{k,n}\|}^2_2 \leq  E_{k,n}(\widetilde{P}_{k,n}) ,
\end{align*}
\begin{align*}
	&{\textrm{C}}^{\textrm{det}}_{4}:P_{c_{k,N}} +\eta_{k,N} \sum_{n=1}^{N}\tau_{4,n}  {\| {\widetilde{\mathbf{{s}}}}_{k,n}\|}^2_2 \leq  E_{k,N}(\widetilde{P}_{k,n}) ,
\end{align*}
where $E_{k,n}(\widetilde{P}_{k,n})$ is defined in \eqref{fg1nl}. We observe that ${\textrm{C}}^{\textrm{det}}_{2}$ is convex. We now consider ${\textrm{C}}^{\textrm{det}}_{3}$ and ${\textrm{C}}^{\textrm{det}}_{4}$. The energy expression ${E}_{k,n} ({\mathbf{x}}_{0})$ is neither convex nor concave w.r.t. ${\mathbf{x}}_{0}$ since $E_{k,n}(\widetilde{P}_{k,n})$ is a non-decreasing concave function w.r.t. $\widetilde{P}_{k,n}$ and $\widetilde{P}_{k,n} ({\mathbf{x}}_{0})$ is a convex function w.r.t. ${\mathbf{x}}_{0}$. Thus, ${\textrm{C}}^{\textrm{det}}_{3}$ and ${\textrm{C}}^{\textrm{det}}_{4}$ are non-convex sets. To deal with this non-convexity, we first define a sufficiently large parameter $\xi_{k,n}$ such that ${\nabla}^{2}_{\mathbf{x}_0} {E}_{k,n} (\mathbf{x}_0) + \xi_{k,n} \mathbf{I}_{{{M}}_h} \succeq \mathbf{0},$ as a convex function plus a concave one:
\begin{equation} \label{jj}
	{E}_{k,n} (\mathbf{x}_0)= \underbrace{ {E}_{k,n} (\mathbf{x}_0) + \frac{1}{2} \xi_{k,n} {\mathbf{x}_0}^{{H}} \mathbf{x}_0}_{\textrm{convex}}
	\underbrace{- \frac{1}{2} \xi_{k,n} {\mathbf{x}_0}^{{H}} \mathbf{x}_0}_{\textrm{concave}}.
\end{equation}
See Appendix~\ref{app4} for such a selection of $\xi_{k,n}$. We now apply the MM technique on ${\textrm{C}}^{\textrm{det}}_{3}$ and ${\textrm{C}}^{\textrm{det}}_{4}$ to obtain convex constraints. To this end, we keep the concave part and minorize the convex part of ${E}_{k,n} ({\mathbf{x}}_{0})$ as follows
\begin{equation}\label{mm}
	{E}_{k,n} \left({\mathbf{x}}^{(\kappa-1)}_{0}\right) + \frac{1}{2} \xi_{k,n} \left({{\mathbf{x}}^{(\kappa-1)}_0}\right)^{{H}} {\mathbf{x}}^{(\kappa-1)}_0 +\Re \left \lbrace {{\mathbf{u}}}^{(\kappa)}_{k,n}   \left( \mathbf{x}_0 - {\mathbf{x}}^{(\kappa-1)}_0 \right) \right \rbrace  -\frac{1}{2} \xi_{k,n} {\mathbf{x}}^{{H}}_0 \mathbf{x}_0,
\end{equation}
where
\begin{align}
	{{\mathbf{u}}}^{(\kappa)}_{k,n} & = \xi_{k,n}   \left({{\mathbf{x}_0}^{(\kappa-1)}}\right)^{{H}} \\ \nonumber & \hspace{2pt}+2 \tau_1 \textrm{exp} \left( c \right) \textrm{exp} \left( a \hspace{1pt} {\left( \textrm{log} \left(   \omega^{(\kappa)}_{k,n}  \right)  \right)}^2  \right) {\left(   \omega^{(\kappa)}_{k,n}  \right) }^{b-1} \left( {2a \textrm{log} \left(   \omega^{(\kappa)}_{k,n}  \right) }    +b \right) \hspace{2pt} \left({{\mathbf{x}}^{(\kappa-1)}_0}\right)^{{H}} \mathbf{H}^{{H}}_{k,n} {\mathbf{H}_{k,n} },
\end{align}
with $\omega^{(\kappa)}_{k,n} = \left({{\mathbf{x}}^{(\kappa-1)}_0}\right)^{{H}} \mathbf{B}_{k,n} \hspace{1pt} {\mathbf{x}}^{(\kappa-1)}_0$. Now, we can rewrite the constraints ${\textrm{C}}^{\textrm{det}}_{3}$ and ${\textrm{C}}^{\textrm{det}}_{4}$ as follows
\begin{align}
	&{\textrm{C}}^{\textrm{det}}_{3}: P_{c_{k,n}} +\eta_{k,n} {\tau_{3,n}}   {\| {\mathbf{{s}}}_{k,n}\|}^2_2 \leq  \eqref{mm} ,
\end{align}
\begin{align}
	&{\textrm{C}}^{\textrm{det}}_{4}:P_{c_{k,N}} +\eta_{k,N} \sum_{n=1}^{N}\tau_{4,n}  {\| {\widetilde{\mathbf{{s}}}}_{k,n}\|}^2_2 \leq   \eqref{mm}.
\end{align}
Finally, to deal with the new design problem, we can modify the algorithm in Table \ref{table:method2} by replacing the constraints ${\textrm{C}}_2$, ${\textrm{C}}^{a}_3$, and ${\textrm{C}}^{a}_4$ in \eqref{maxmin1001} with ${\textrm{C}}^{\textrm{det}}_{2}-{\textrm{C}}^{\textrm{det}}_{4}$ above.
\subsection{Sum Throughput Maximization} \label{sum}
The proposed max-min throughput method can be straightforwardly extended to the sum throughput maximization problem. Considering the sum throughput of the network. i.e., $\sum_{k=1}^{K} \sum_{n=1}^{N}$ $R_{k,n}^{(\textrm{U})}$, we cast the following sum throughput optimization problem
\begin{eqnarray}\label{sum1}
\max_{\hm{\tau}  ,{\mathbf{{Q}}}, {\mathbf{{S}}}, {\widetilde{\mathbf{{S}}}} } &&  \sum_{k=1}^{K} \sum_{n=1}^{N}  R_{k,n}^{(\textrm{U})}\\ \nonumber
\mbox{s.t.}\;\;&& \textrm{C}_{1} - \textrm{C}_{4}.
\end{eqnarray}
Similar to the problem in \eqref{maxmin1}, the problem in \eqref{sum1} is non-convex due to the coupled design variables in the objective function and in the constraints $\textrm{C}_{3}$ and $\textrm{C}_{4}$. Therefore, we employ AO with partitioning $[\hm{\tau} \hspace{5pt} \vdots \hspace{5pt} \mathbf{{Q}}, \mathbf{{S}}, {\widetilde{\mathbf{{S}}}}]$. The resulting sub-problem w.r.t. the parameter $\hm{\tau}$ is an LP and can  be solved via e.g. interior point methods (similar to the Subsection \ref{keyLP}). Then, by introducing the auxiliary variables $\phi_{k,n}, \hspace{2pt} \forall k, n \neq N$ and using \eqref{fg11} and \eqref{fg1101}, the Problem \eqref{sum1} w.r.t. $[\mathbf{{Q}}, \mathbf{{S}}, {\widetilde{\mathbf{{S}}}}]$ can be recast as
\begin{eqnarray}\label{sum2}
\max_{   {\mathbf{{Q}}}, {\mathbf{{S}}},\widetilde{\mathbf{{S}}},\hm{\Phi}} & & \sum_{k=1}^{{K}} \left \lbrace {R}^{(\textrm{CH-HAP})}_{k,N} \left( \widetilde{\mathbf{{S}}} \right)  + \sum_{n=1}^{{N-1}} {\phi_{k,n}} \right \rbrace \\ \nonumber
\mbox{s.t.}\;\;&& \textrm{C}^{\textrm{sum}}_{5,1}: R_{k,n}^{(\textrm{U-CH})} \left( \mathbf{{S}} \right)   \geq {\phi_{k,n}}, \hspace{5pt} \textrm{C}^{\textrm{sum}}_{5,2}: R_{k,n}^{(\textrm{U-HAP})} \left( {\mathbf{{S}}} \right) +R_{k,n}^{(\textrm{CH-HAP})}\left( \widetilde{\mathbf{{S}}} \right)  \geq  {\phi_{k,n}}, \hspace{2pt} \forall k,n \neq N, \\ \nonumber \;\;&&
\textrm{C}_{2} - \textrm{C}_{4},
\end{eqnarray}
where $\hm{\Phi}=[\phi_{k,n}, \hspace{2pt} \forall k, n \neq N]$. The constraints $\textrm{C}^{\textrm{sum}}_{5,1}$ and $\textrm{C}^{\textrm{sum}}_{5,2}$ can be tackled similar to the constraints $\textrm{C}^{a}_{5,1}$ and $\textrm{C}^{a}_{5,2}$ in Subsection \ref{keymat}. Moreover, similar to the minorization process in \eqref{fg60}-\eqref{fg6c}, the objective function of this problem (considering SIC) can be dealt with by neglecting the constant terms. Consequently, we can handle the sum throughput maximization problem in \eqref{sum2} at the $\kappa$th MM iteration by:
\begin{eqnarray}\label{sum3}
\min_{ \mathbf{Q} , \mathbf{s} , \widetilde{\mathbf{s}} ,\hm{\Phi} } & & \sum_{k=1}^{{K}} \left \lbrace \sum_{j=k}^{{K}}  {{\widetilde{\mathbf{s}}}^{{H}}_{j,N}} \bar{\mathbf{\Upsilon}}^{(\kappa-1)}_{j,N} {\widetilde{\mathbf{s}}}_{j,N}
+2  \Re \left\lbrace \left(\bar{\mathbf{{v}}}^{(\kappa-1)}_{k,N} \right)^{{H}} { \widetilde{\mathbf{s}}}_{k,N} \right\rbrace  - \sum_{n=1}^{N-1} \phi_{k,n} \right \rbrace \\ \nonumber
\mbox{s.t.}\;\;&& \textrm{C}^{\textrm{sum}}_{5,1}: \eqref{fg6001} \geq \phi_{k,n} , \hspace{10pt}  \textrm{C}^{\textrm{sum}}_{5,2}: \eqref{fg60101} \geq \phi_{k,n}, \hspace{2pt} \forall k, 1 \leq n \leq N-1,\\ \nonumber
&& \textrm{C}_{2}, \hspace{4pt} \textrm{C}^{a}_{3}, \hspace{4pt} \textrm{C}^{a}_{4}.
\end{eqnarray}
We see that the problem in \eqref{sum3} is convex, i.e., a QCQP w.r.t. $[{\mathbf{s}}, \widetilde{\mathbf{s}}]$ with linear constraints on matrix $\mathbf{{Q}}$ and $\hm{\Phi}$. Thus, this problem can be solved efficiently by using interior point methods.
\section{Numerical Examples}\label{num}
\begin{figure}
	\centering
	\begin{tikzpicture}[even odd rule,rounded corners=2pt,x=12pt,y=12pt,scale=.7]
	\draw[dashed,black!100] (3.5,2.5)--+(14.5,4.5);
	\draw[dashed,black!100] (3.5,2.5)--+(14.5,-4.5);
	\draw[dashed,black!100] (3.5,2.5)--+(10.5,10.5);
	\draw[dashed,black!100] (3.5,2.5)--+(10.8,-10.8);
	\draw[thick,fill=red!10] (-.5,0) rectangle (3.3,5) node[midway]{ HAP};
	\draw[dashed,black!100] (10.5,-5.5) arc (-50:50:10);
	\draw[dashed,black!100] (13.5,-9.5) arc (-50:48:15);
	
	\draw[thick,black!100] (5.4,1.9) arc (-18:18:2) ;
	\draw[thick,black!100] (5.4,.6) arc (-41:-12:2.5);
	\draw[thick,black!100] (5.8,3.2) arc (20:48:2.5);
	\node [] at (6.2,2.5) {\footnotesize $\theta$};
	\node [] at (6.2,4) {\footnotesize $\theta$};
	\node [] at (6.2,.8) {\footnotesize $\theta$};
	
	\draw[thick,fill=blue!20] (9.5,8.7) rectangle ++(2,2) node[midway]{ \footnotesize $U_{1}^{2}$};
	\draw[thick,fill=green!10] (13.9,12.8) rectangle ++(2,2) node[midway]{\footnotesize $U_{1}^{1}$};
	
	\draw[thick,fill=blue!20] (12.4,4.5) rectangle ++(2,2) node[midway]{\footnotesize $U_{2}^{2}$};
	\draw[thick,fill=green!10] (17.9,6.8) rectangle ++(2,2) node[midway]{\footnotesize $U_{2}^{1}$};

	\draw[thick,fill=blue!20] (12.4,-1.9) rectangle ++(2,2) node[midway]{ \footnotesize$U_{3}^{2}$};
	\draw[thick,fill=green!10] (18.2,-4) rectangle ++(2,2) node[midway]{\footnotesize $U_{3}^{1}$};
	
	\draw[thick,fill=blue!20] (9.9,-5.9) rectangle ++(2,2) node[midway]{\footnotesize $U_{4}^{2}$};
	\draw[thick,fill=green!10] (12.7,-10.3) rectangle ++(2,2) node[midway]{\footnotesize $U_{4}^{1}$};

	\draw[<->,thick,black!100] (13.4,6.7) --node[right] {\footnotesize $d^{(\textrm{U-CH})}_{2,1,1}$} (14.7,6.2+6.5) ;
	\draw[<->,thick,black!100] (14.7+.2-.5,5.8-1.3) --node[right] {\footnotesize $d^{(\textrm{U-CH})}_{2,3,1}$} (14.7+3.4,5.8-7.7);
	\draw[<->,thick,black!100] (14.7+3.4,5.8-8.2) --node[below,sloped] {\footnotesize $d^{(\textrm{U-CH})}_{3,3,1}$} (14.5,5.8-7.15);
	\draw[<->,thick,black!100] (3.6,2.4) -- node[below,sloped] {\footnotesize $d^{(\textrm{U-HAP})}_{4,2}$}(3.6+6.3,2.4-6.3);
	\draw[<->,thick,black!100]  (3.4,3)-- node[above,sloped] {\footnotesize $d^{(\textrm{U-HAP})}_{1,1}$} (3.6+10.2,4+9.4) ;
	\draw[thick,fill=blue!20] (9.5,8.7) rectangle ++(2,2) node[midway]{\footnotesize  $U_{1}^{2}$};
	\end{tikzpicture}
	\caption{The geometry for the simulation setup for the case of $N=2$.}
	\label{ht1550}
	\centering
\end{figure}
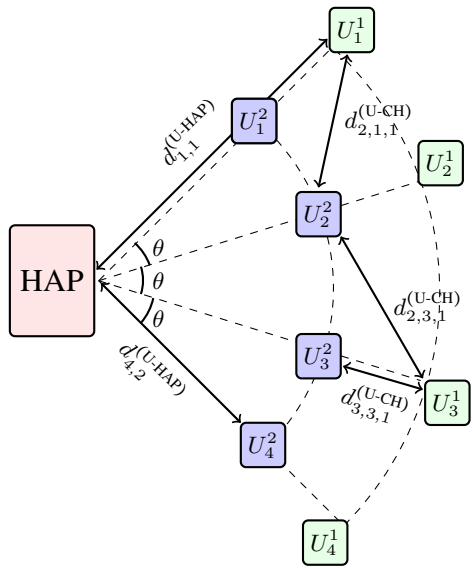

In this section, we evaluate  the proposed method in different scenarios via Monte-Carlo simulations. The channels from HAP to users and the channels between user $U_{i}^{n}$ and $k$th CH $U_{k}^{N}$ are modeled as $\mathbf{H}_{k,n}= {0.1 \left(\frac{d^{(\textrm{U-HAP})}_{k,n}}{d_0} \right)^{\frac{-\alpha}{2}}} \widetilde{\mathbf{H}}_{k,n} ,\forall k,n$ and ${\mathbf{G}}_{k,n,i}={0.1 \left(\frac{d^{(\textrm{U-CH})}_{k,n,i}}{d_0} \right)^{\frac{-\alpha}{2}}} \widetilde{\mathbf{G}}_{k,n,i}, \forall k,i, n \neq N$, respectively, where $d_0=1 \hspace{2pt}\textrm{meters (m)}$ is a reference distance, $d^{(\textrm{U-HAP})}_{k,n}$ is the distance between HAP and users, ${d^{(\textrm{U-CH})}_{k,n,i}}$ is the distance between user $U_{i}^{n}$ and $k$th CH (see Fig. \ref{ht1550}) and $\alpha$ is the path-loss exponent. We assume that the entries of $\widetilde{\mathbf{H}}_{k,n}$ and $\widetilde{\mathbf{G}}_{k,n,i}$ are i.i.d. circularly symmetric complex Gaussian random variables with zero mean and unit variance. We set $N=2$ and also $2d^{(\textrm{U-HAP})}_{k,2}= 2d^{(\textrm{U-CH})}_{k,1,k} = d^{(\textrm{U-HAP})}_{k,1} = 10\hspace{1pt}\textrm{m}, \forall k,n,$ $\theta=30^{\circ}$ (as shown in Fig. \ref{ht1550}) and $\alpha=3$, unless otherwise stated. We also consider HAP and the users with ${M}_{h}={M}_{k,n}={M}=4,\forall k,n$ antennas and set $p_0=3\hspace{2.5pt}\textrm{watts (w)}$ unless otherwise specified. We assume the receiver noise vectors to be white with ${\mathbf{L}}^{(\textrm{U-HAP})}_{n}={\mathbf{L}}^{(\textrm{U-CH})}_{k,n}= \sigma^{2} \mathbf{I}_{{M}}, \forall k, n \neq N$ and $\sigma^{2}= -80\hspace{2pt}\textrm{dBm}$. We further assume ${K}=4$, $P_{c_{k,n}}=-23\hspace{2pt}\textrm{dBm}, \forall k,n$, and power amplifier inefficiency factor $\eta_{k,n}=1, \forall k,n$. Moreover, we use a bandwidth of $1\hspace{2pt}\textrm{MHz}$
and the curve fitting parameters $a=-0.0977$, $b=-0.9151$ and $c=-11.1648$ as suggested in \cite[Fig. 5.a]{clerckx2018beneficial}. Finally, we solve the convex optimization problems using CVX package \cite{cvx} for ${\rho}_{{{h}_{k,n}}} = {\rho}_{{{g}_{k,n,i}}}={\rho}=0.95, \forall k,n,i$.
\subsection{Convergence of the Proposed Algorithm}
 \begin{figure}
 \includegraphics[scale=.5]{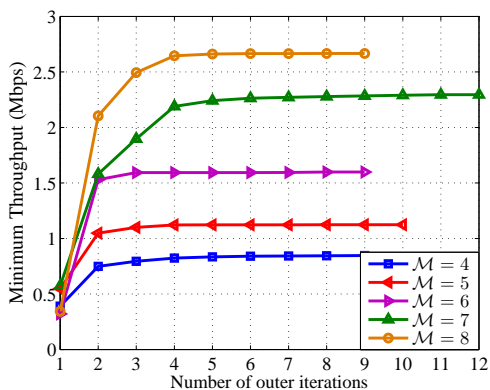}
  \centering
 \caption{The values of the objective function in \eqref{maxmin1} versus the number of outer iterations.}
 \label{PA4}
 \end{figure}
Fig. \ref{PA4} shows the values of the max-min throughput, i.e., the objective values in \eqref{maxmin1} versus the number of outer iterations (see Table \ref{table:method2}), for different number of antennas at the HAP and the users (${M}$). As expected, the max-min throughput increases at each iteration and quickly converges within small number of iterations.
\subsection{The Effect of the CH Locations}
\begin{figure}
	\hfill
	\subfigure[]{\includegraphics[scale=.492]{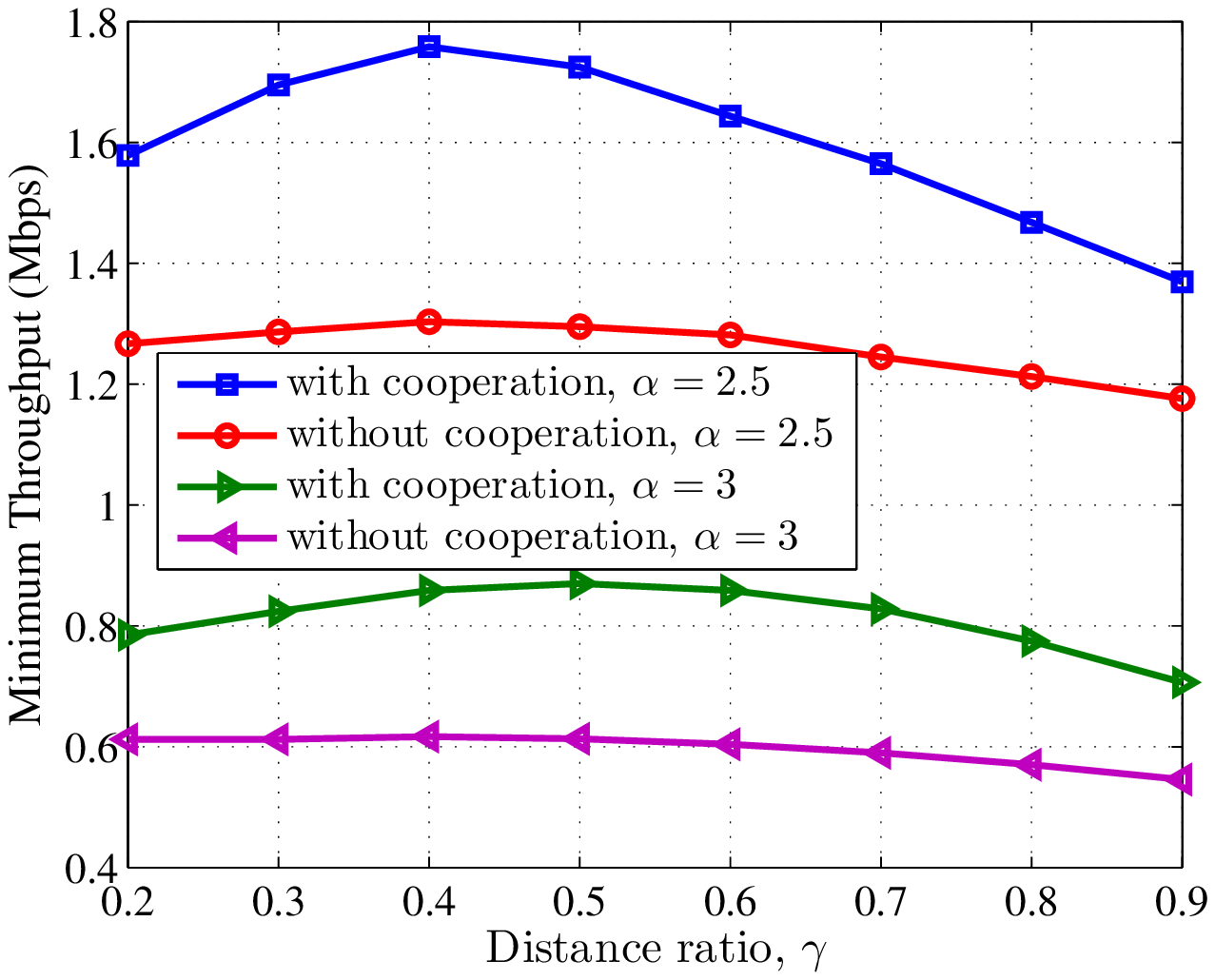}}
	\hfill
	\subfigure[]{\includegraphics[scale=.49]{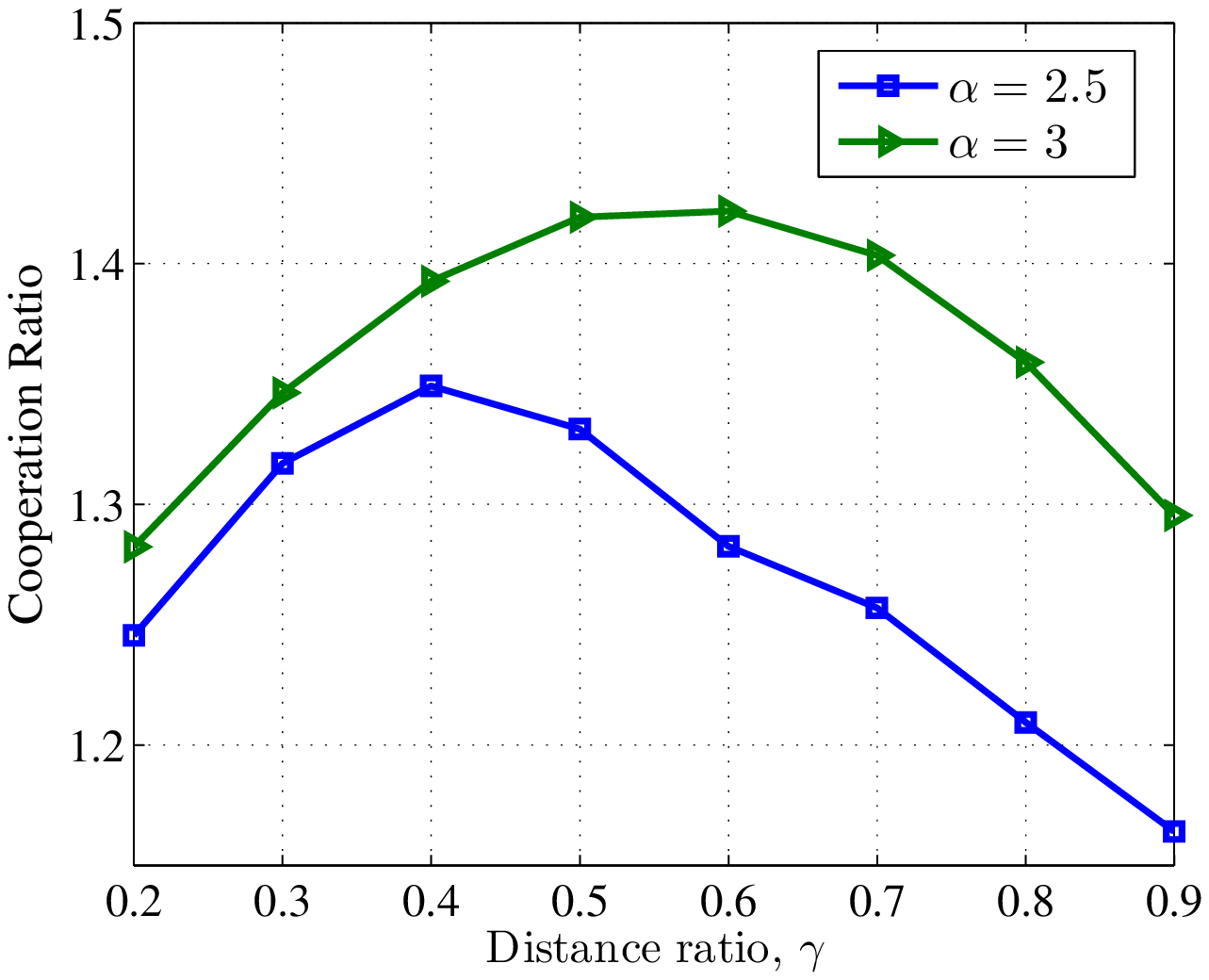}}
	\hfill
	\caption{Comparison of the cooperative and non-cooperative methods versus various distance ratio: (a) minimum throughput; cooperative method: $\gamma_{\textrm{opt}}=0.4$ for $\alpha=2.5$ and $\gamma_{\textrm{opt}}=0.5$ for $\alpha=3$, non-cooperative method: $\gamma_{\textrm{opt}}=0.4$ for both $\alpha=2.5,3$, (b) cooperation ratio; $\gamma_{\textrm{opt}}=0.4$ for $\alpha=2.5$ and $\gamma_{\textrm{opt}}=0.6$ for $\alpha=3$.}
	\label{PA10}
\end{figure}
In this scenario, we study the effect of the distance between HAP, $k$th CH, i.e. $U^{2}_k$, and $U^{1}_k$ on the max-min throughput performance. For this purpose, we define $\gamma=\frac{d^{(\textrm{U-HAP})}_{k,2}}{d^{(\textrm{U-HAP})}_{k,1}}$ as the distance ratio as shown in Fig. \ref{ht1550}. The max-min throughput versus $\gamma$ is plotted in Fig. \ref{PA10}.a for different values of $\alpha$. In this figure, we also compare the proposed cooperative resource allocation method with a non-cooperative method. In the non-cooperative method, CHs do not perform their relaying duty in the fourth phase; indeed, for the time duration of the fourth phase we can write $\tau_{4,n}=0, \hspace{2pt}\forall n \neq N$ and therefore, $\tau_4 = \tau_{4,N}$ (see Fig. \ref{h1t}). We observe that the cooperation leads to a significant increase for the achievable minimum throughput, for all values of $\gamma$ and $\alpha$. Also, the minimum throughput of the method with cooperation first increases until $\gamma \leq \gamma_{\textrm{opt}}$, and then decreases by increasing $\gamma$, i.e., when $\gamma \geq \gamma_{\textrm{opt}}$. Because the doubly near-far problem is more severe for larger values of $\alpha$, the value of $\gamma_{\textrm{opt}}$ increases w.r.t. $\alpha$ for the method with cooperation. In light of Fig.~\ref{PA10}.a, we define a cooperation ratio as $\frac{\textrm{coopreration}}{\textrm{non-coopreration}}$ and plot it in Fig. \ref{PA10}.b versus distance ratio. The optimum location of CHs associated with the largest cooperation ratio can be observed from this figure.
\subsection{Minimum Throughput Versus Maximum Power of HAP}
\begin{figure}
	\centering
	\includegraphics[scale=.6]{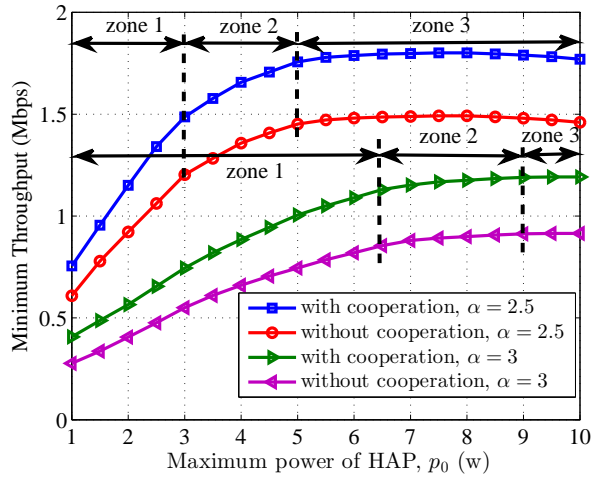}
	\caption{Minimum throughput versus maximum power budget at HAP.}
	\label{PA0}
	\centering
\end{figure}

In Fig. \ref{PA0}, we plot the max-min throughput as a function of the maximum power budget at HAP i.e., $p_0$, for a constant distance ratio $\gamma=0.8$ (see Fig.~\ref{ht1550}) for different values of the path-loss exponent $\alpha$ and comparing the cooperative and non-cooperative methods. We observe that the proposed cooperative resource allocation method achieves higher throughput than the non-cooperative method. Moreover, the curves in Fig.~\ref{PA0}, can be interpreted in three zones. In zone 1, the EH circuits of both $k$th CH and $U_k^1$ in each cluster operate in their linear region. In zone 2, the EH circuit of $U_k^1$ operates in its linear region and the EH circuit of $k$th CH is saturated. In zone 3, the EH circuits of both are saturated.
Note that minimum throughput gain from the cooperation for $\alpha=3$ is less than that of $\alpha=2.5$ in zone 1 (linear zone) due to the larger channel attenuation.
\subsection{The Max-min Versus Sum Throughput Optimizations}
\begin{figure}
\centering
\includegraphics[scale=.6]{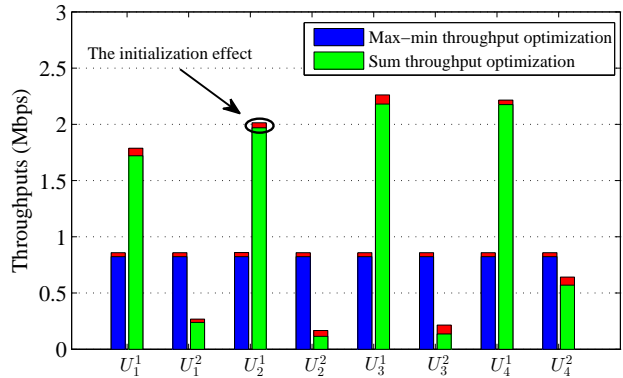}
\caption{The user throughputs in the max-min and sum throughput optimizations.}
\label{PA5}
\centering
\end{figure}

Fig. \ref{PA5} compares the throughput obtained from the max-min and sum throughput optimization problems, for the proposed (cooperative) method. It is observed that in sum throughput optimization, the values of the minimum throughput and the sum throughput of all pairs are equal to $\min_{\substack{1 \leq k \leq 4 \\ 1 \leq n \leq 2}} {R}^{(\textrm{U})}_{k,n} =0.12$ Mbps and $\sum_{k=1}^{4} \sum_{n=1}^{2} R^{(\textrm{U})}_{k,n} =9.11$ Mbps, respectively; whereas for the max-min case we have $\min_{\substack{1 \leq k \leq 4 \\ 1 \leq n \leq 2}} {R}^{(\textrm{U})}_{k,n} =0.83$ Mbps and $\sum_{k=1}^{4} \sum_{n=1}^{2} R^{(\textrm{U})}_{k,n} =6.64$ Mbps. Therefore, we can conclude that the max-min throughput method achieves fairness among users by compromising sum throughput. {Note that the objective in the sum throughput maximization method is to maximize the sum throughput of all users, and as a result, we numerically observed that the time allocated to the time slots $\tau_{4,n}, \forall n \neq N$, i.e., relaying time slots of the sum throughput problem is very small.}.

We illustrate the effect of the initialization in step 0 of Table~\ref{table:method2} on the performance of the both methods. Considering red parts, it can be seen that the quality of the obtained solution has a small sensitivity w.r.t. various initializations. Precisely, for 10 different initial points, the maximum level of relative error (i.e., the maximum deviation from average value divided by average value) for the minimum throughput of the max-min problem and sum throughput of the sum throughput maximization problem are respectively $2.37\%$ and $3.32\%$.
\subsection{The Effect of Inter-Cluster Interference }
\begin{figure}[b]
\centering
\includegraphics[scale=.48]{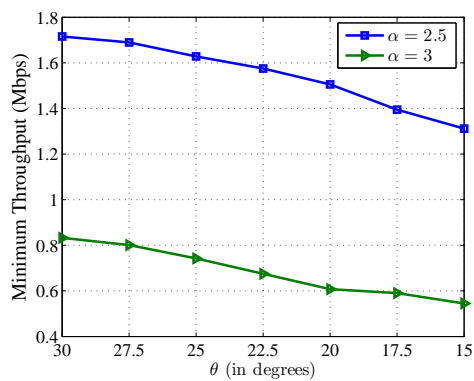}
\caption{Minimum throughput versus the parameter $\theta$ (in degrees).}
\label{tetta}
\centering
\end{figure}

Fig. \ref{tetta} illustrates the minimum throughput of the cooperative method versus the parameter $\theta$ (see Fig. \ref{ht1550}). We observe that as $\theta$ decreases, the minimum throughput decreases for both scenarios with $\alpha=2.5$ and $\alpha=3$. This is due to the fact that as $\theta$ increases, the inter-cluster interference increases which leads to a reduction in $R^{(\textrm{U-CH})}_{k,n}, \forall k, n \neq N$ and thus a reduction in $R_{k,n}^{(\textrm{U})}, \forall k, n \neq N$ (see \eqref{fg11}). Note that this throughput reduction for $\alpha=3$ is less than the case of $\alpha=2.5$, because there is more inter-user interference for $\alpha=2.5$.
\subsection{The Effect of Channel Estimation Error}
\begin{figure}
	\hfill
	\subfigure[]{\includegraphics[scale=.492]{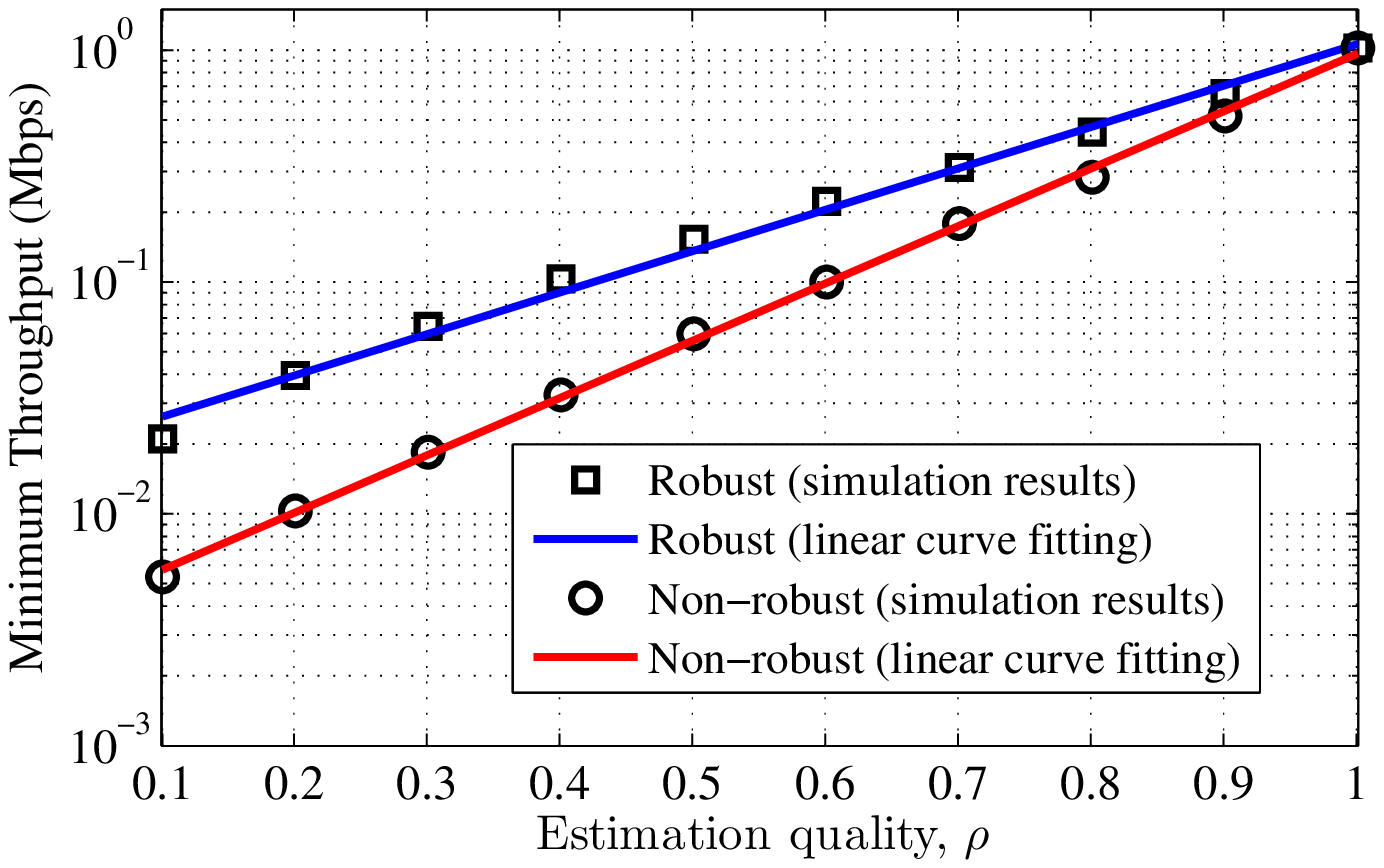}}
	\hfill
	\subfigure[]{\includegraphics[scale=.49]{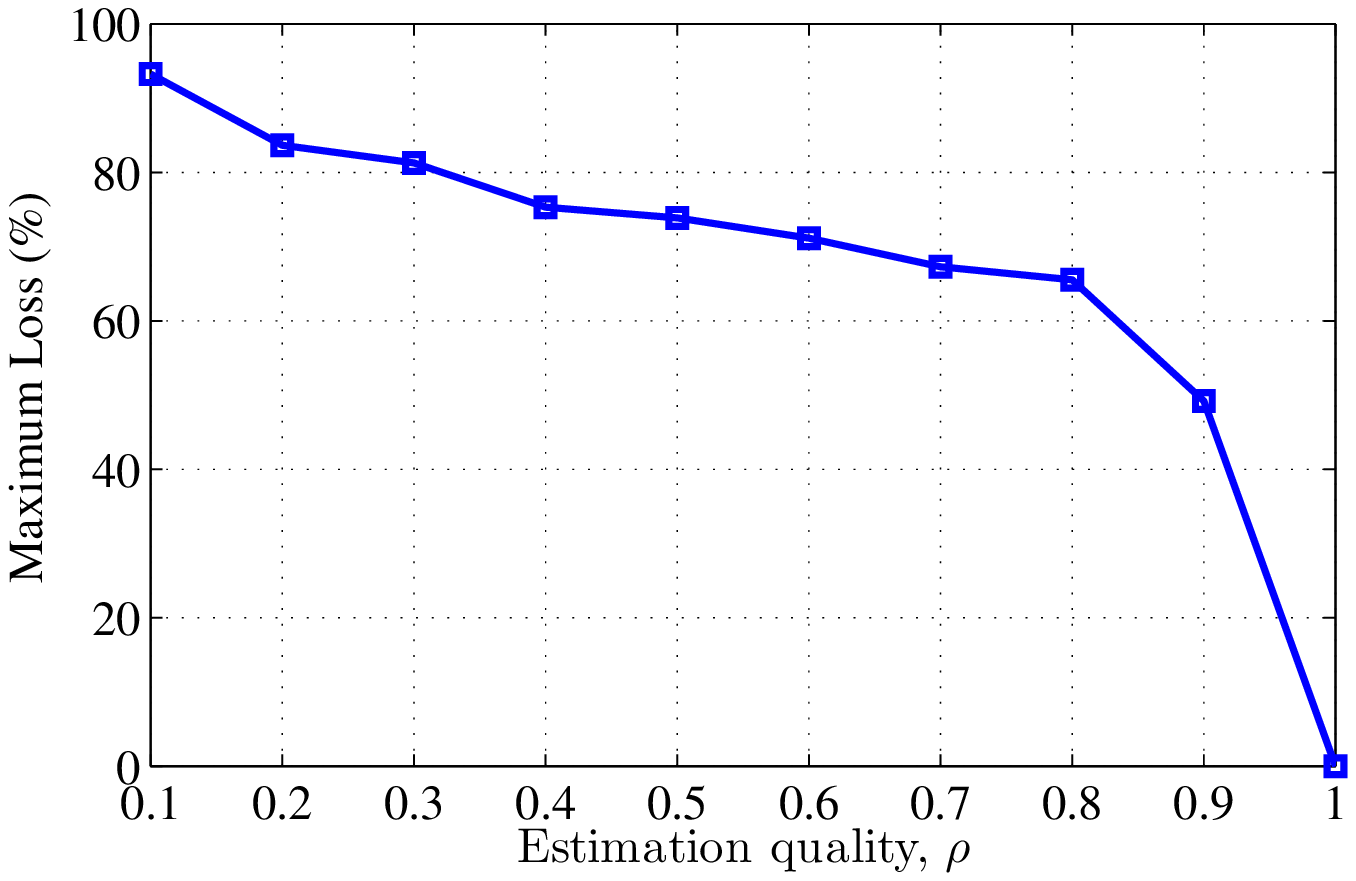}}
	\hfill
	\caption{The effect of channel estimation error: (a) linear curve fitting (in log-scale) for minimum throughput versus the estimation quality $\rho$; proposed (robust) method: $a_{\rho}=4.1123,\hspace{2pt}b_{\rho}=7.4633$, non-robust method: $a_{\rho}=5.6956,\hspace{2pt}b_{\rho}=5.7844$, (b) maximum value of the loss parameter $\chi (\rho)$ versus the estimation quality $\rho$.}
	\label{PA6}
\end{figure}

In this subsection, we study the effect of channel estimation error by comparing the robust and the non-robust schemes for the cooperative max-min method in Fig. \ref{PA6}. The robust scheme considers the errors of channel estimation in the resource allocation stage as opposed to the non-robust scheme. In this case, we apply a linear curve fitting to both methods in a least-squares sense in a log-scale as follows
\begin{equation}
{R}^{(\textrm{U})}_{dB}=\textrm{log} \left( \min_{k,n} {R}^{(\textrm{U})}_{k,n}  \right) =a_{\rho} \rho + b_{\rho}
\end{equation}
where $a_{\rho}$ and $b_{\rho}$ are the coefficients of the polynomial.
We observe in  Fig. \ref{PA6}.a that the minimum throughput increases as the estimation quality $\rho$ increases.
 Note that in Fig. \ref{PA6}.a, we consider the performance evaluation of the robust and the non-robust schemes in the average sense. However, in some cases, the non-robust scheme has a much worse performance than the robust one. Therefore, we define the loss parameter as
\begin{equation}
\chi (\rho) = 1-\frac{{R}^{(\textrm{nr})}(\rho)}{{R}^{(\textrm{r})}(\rho)},
\end{equation}
where ${R}^{(\textrm{nr})}$ and ${R}^{(\textrm{r})}$ represent the minimum throughput for the robust and the non-robust schemes, respectively. Fig. \ref{PA6}.b shows the maximum of the loss parameter over 100 realizations of the channel matrices. As expected, the loss decreases as the estimation quality increases, and tends to zero for perfect CSI as $\rho \rightarrow 1$. We also observe that the proposed (robust) scheme provides enhanced throughput compared to the non-robust one even for relatively large values of the estimation quality.
\subsection{The Impact of CH Selection for $N > 2$}
\begin{figure}
\centering
\includegraphics[scale=.48]{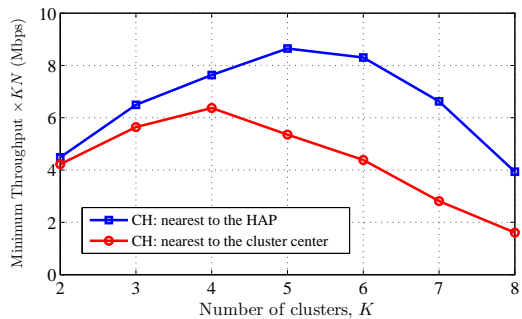}
\caption{Minimum throughput$\hspace{1pt}\times KN$ versus the number of clusters $K$ for the two CH selection methods.}
\label{httttttt0}
\centering
\end{figure}

Herein, we aim to analyze the impact of the CH assignment to the minimum throughput performance for the cooperative max-min throughput optimization problem. In the previous subsections for $N=2$, it was trivial to select the user which is closest to the HAP as CH in each cluster.
In this scenario, we assume that the users are uniformly distributed
in each cluster within a circle with a radius of $5\hspace{1.5pt}\textrm{m}$. Furthermore, the center of the clusters are set to be $10\hspace{1.5pt}\textrm{m}$ apart and also  $10\hspace{1.5pt}\textrm{m}$ away from the HAP. Fig.~\ref{httttttt0} compares the performance of the two CH selection methods in which the nearest user to the HAP$\slash$cluster center is selected as CH for each cluster. Precisely, Fig.~\ref{httttttt0} shows the total minimum throughput, i.e., minimum throughput$\hspace{1pt}\times KN$ of two mentioned methods versus number of clusters $K$ for the case of $N=4$. We observe a superior performance when the CH is the nearest user to the cluster center as compared to the one nearest to HAP in Fig.~\ref{httttttt0}. This is because the maximum distance between users and CH in each cluster is larger and as a result the minimum throughput will be lower.
\section{Conclusion} \label{con}
In this paper, we proposed a fair cooperative multi-cluster method in a WPCN where both the HAP and users have multiple antennas. In the downlink phase of the network, the HAP transfers RF energy toward the users and then, the users, via the harvested energy in the aforementioned phase, transmit their information to the HAP with cooperation in the uplink phase. We considered the minimum and sum throughputs of the users as the design metrics and cast the optimization problem w.r.t. the time allocation vector, EB matrix, and covariance matrices of the users. The design problems were non-convex and so hard to solve. Therefore, we resort to the AO approach to deal with the problems. The resulting sub-problems w.r.t. the time allocation vector were convex; however, the sub-problems w.r.t. the EB and covariance matrices were non-convex and hence, the MM technique was employed to deal with them. We further considered the imperfect CSI and non-linear EH circuit in our design methodology. Various numerical examples illustrated the effectiveness of the proposed cooperative max-min scheme to achieve fairness among users. Studying distributed algorithm for cooperative multi-cluster WPCN to reduce signaling overhead especially in systems with large number of users can be a topic of future work.

\appendices
\section{Proof of the LMMSE decoders in \eqref{w1}} \label{app1}
Without loss of generality, we assume that the information stream of the $U^{n}_{k}, 1 \leq n \leq N-1$ and the received signal at the $k$th CH (i.e., $U^{N}_{k}$) in \eqref{fg60n} satisfy $\mathbb{E} \left [ \mathbf{m}_{k,n} \right ]=\mathbb{E} \left [ \mathbf{y}^{(\textrm{U-CH)}}_{k,n} \right ] =0$. Then, for the LMMSE decoder $\mathbf{W}_{k,n}$ we have \cite{kay1993fundamentals1}
\begin{align}
&\mathbf{W}_{k,n} =  \mathbb{E} \left [ \mathbf{m}_{k,n} \left({\mathbf{y}^{(\textrm{U-CH)}}_{k,n}}\right)^{{H}} \right] \left( \mathbb{E} \left[ {\mathbf{y}^{(\textrm{U-CH)}}_{k,n}} \left({\mathbf{y}^{(\textrm{U-CH)}}_{k,n}}\right)^{{H}}  \right] \right) ^{-1} \\ \nonumber
&={\mathbf{{V}}}^{{H}}_{k,n}  {\widehat{ \mathbf{G}}^{{H}}_{k,n,k}} \left ( {{\mathbf{L}}}^{(\textrm{U-CH)}}_{k,n}  + \sum_{j=1}^{{K}} \widehat{ \mathbf{G}}_{k,n,j} {\mathbf{{S}}}_{j,n} \widehat{ \mathbf{G}}^{{H}}_{k,n,j} + \mathbb{E} \left [ \sum_{j=1}^{{K}} \sum_{j^{\prime} =1}^{{K}} {\Delta \mathbf{G}}_{k,n,j} {\mathbf{{V}}}_{j,n}   \mathbf{m}_{j,n} \mathbf{m}^{{H}}_{j^{\prime},n} {\mathbf{{V}}}^{{H}}_{j^{\prime},n} \Delta \mathbf{G}^{{H}}_{k,n,j^{\prime}} \right] \right) ^{-1}
\\ \nonumber
&={\mathbf{{V}}}^{{H}}_{k,n}   {\widehat{ \mathbf{G}}^{{H}}_{k,n,k}} \left( {{\mathbf{L}}}^{(\textrm{U-CH)}}_{k,n}  + \sum_{j=1}^{{K}} \widehat{ \mathbf{G}}_{k,n,j} {\mathbf{{S}}}_{j,n} \widehat{ \mathbf{G}}^{{H}}_{k,n,j} + \sum_{j=1}^{{K}} \sigma^{2}_{{g,\Delta}_{k,n,j}}  \textrm{tr}  \left \lbrace  {\mathbf{{S}}}_{j,n} \right \rbrace \mathbf{I}_{{M}_{k,N}}  \right) ^{-1}.
\end{align}	
\section{The derivation of the throughput expression in \eqref{fg60}} \label{appnew}
Let denote $\mathbf{D}^{-1}_{k,n}$ as inversion of the matrix $\mathbf{D}_{k,n}$, with
\begin{equation}
\mathbf{{D}}^{-1}_{k,n} =
\begin{bmatrix}
\mathbf{{D}}^{-1}_{k,n,11} \in \complexC^{{M}_{k,n} \times {M}_{k,n} }
\hspace*{2mm}&\hspace*{2mm}   \mathbf{{D}}^{-1}_{k,n,12} \in \complexC^{{M}_{k,n} \times {M}_{k,N} }  \vspace*{3mm} \\
\mathbf{{D}}^{-1}_{k,n,21} \in \complexC^{{M}_{k,N} \times {M}_{k,n} } \hspace*{2mm}&\hspace*{2mm}  \mathbf{{D}}^{-1}_{k,n,22} \in \complexC^{{M}_{k,N} \times {M}_{k,N} }
\end{bmatrix}, \hspace{5pt} \forall k,1 \leq n \leq N-1.
\end{equation}
By using blockwise matrix inversion lemma (see, e.g., \cite[Appendix A]{stoica2005spectral}), we can write:
\begin{align}
 \mathbf{{D}}^{-1}_{k,n,11}=&
 \Bigg(\mathbf{I}_{{M}_{k,n}} - \left({\mathbf{{S}}}^{\frac{1}{2}}_{k,n}\right)^{{H}}  \widehat{ \mathbf{G}}^{{H}}_{k,n,k} \bigg({{\mathbf{L}}}_{k,n}^{(\textrm{U-CH})}  +\sum_{j=1}^{{K}}               \widehat{ \mathbf{G}}_{k,n,j} {\mathbf{{S}}}_{j,n}  \widehat{ \mathbf{G}}^{{H}}_{k,n,j}  \\ \nonumber & \hspace{10pt}+  \sum_{j=1}^{{K}} \sigma^{2}_{{g,\Delta}_{k,n,j}} \textrm{tr}  \left \lbrace  {\mathbf{{S}}}_{j,n} \right \rbrace \mathbf{I}_{{M}_{k,N}} \bigg) ^{-1} \widehat{ \mathbf{G}}_{k,n,k} {\mathbf{{S}}}^{\frac{1}{2}}_{k,n}  \Bigg) ^{-1}, \hspace{3pt} \forall k, 1 \leq n \leq N-1.
\end{align}
Then, by using Woodbury matrix identity, i.e., ${(\mathbf{F} + \mathbf{U} \mathbf{C} \mathbf{V})}^{-1} = \mathbf{F}^{-1} - \mathbf{F}^{-1} \mathbf{U} {(\mathbf{C}^{-1} + \mathbf{V} \mathbf{F}^{-1} \mathbf{U})}^{-1} \mathbf{V} \mathbf{F}^{-1}$,
we can rewrite $\mathbf{{D}}^{-1}_{k,n,11}$ as
\begin{align} \label{keyl}
\mathbf{{D}}^{-1}_{k,n,11}=&
\mathbf{I}_{{M}_{k,n}} + \left({\mathbf{{S}}}^{\frac{1}{2}}_{k,n}\right)^{{H}}  \widehat{ \mathbf{G}}^{{H}}_{k,n,k} \bigg({{\mathbf{L}}}_{k,n}^{(\textrm{U-CH})}  +\sum_{j=1,j \neq k}^{{K}}               \widehat{ \mathbf{G}}_{k,n,j} {\mathbf{{S}}}_{j,n}  \widehat{ \mathbf{G}}^{{H}}_{k,n,j}  \\ \nonumber & +  \sum_{j=1}^{{K}} \sigma^{2}_{{g,\Delta}_{k,n,j}} \textrm{tr}  \left \lbrace  {\mathbf{{S}}}_{j,n} \right \rbrace \mathbf{I}_{{M}_{k,N}} \bigg) ^{-1} \widehat{ \mathbf{G}}_{k,n,k} {\mathbf{{S}}}^{\frac{1}{2}}_{k,n}   , \hspace{3pt} \forall k, 1 \leq n \leq N-1.
\end{align}
Finally, by substituting \eqref{keyl} in \eqref{fg60} and using Sylvester’s determinant property i.e., $\textrm{det}(\mathbf{I} + \mathbf{AB}) =
\textrm{det}(\mathbf{I}+\mathbf{BA})$, \eqref{fg6} is obtained.

\section{Details on variables in \eqref{fg60101} and \eqref{fg60111}} \label{app3}
Note that variables that are used for the constraints $\textrm{C}^{a}_{5,2}$ and $\textrm{C}^{a}_{6}$ of the Problem \eqref{maxmin101} in \eqref{fg60101} and \eqref{fg60111} can be written as

\begin{equation*}
 \widetilde{\mathbf{A}}_{k,n}=[\mathbf{I}_{{M}_{k,n}}, \mathbf{0}_{{M}_{k,n} \times {M}_h}]^{T}, \forall k, 1 \leq n \leq N-1, \hspace{10pt}  \bar{\mathbf{A}}_{k,n}=[\mathbf{I}_{{M}_{k,N}}, \mathbf{0}_{{M}_{k,N} \times {M}_h}]^{T}, \forall k,n,
\label{fg614}
\end{equation*}
\begin{equation*} \label{201}
\widetilde{\mathbf{{D}}}_{k,n} =
\begin{bmatrix}
\mathbf{I}_{{M}_{k,n}}  \hspace*{.5mm}&\hspace*{.5mm}   \left({\mathbf{{S}}}^{\frac{1}{2}}_{k,n}  \right) ^{{H}} \widehat{\mathbf{{H}}}_{k,n} \vspace*{2mm} \\
\widehat{\mathbf{H}}^{{H}}_{k,n}  {{\mathbf{{S}}}^{\frac{1}{2}}_{k,n}} \hspace*{.5mm}&\hspace*{.5mm} {{\mathbf{L}}}_n^{(\textrm{U-HAP})} +\sum_{j=k}^{{K}}  \left \lbrace {\widehat{\mathbf{{H}}}^{{H}}_{j,n}}   {{\mathbf{{S}}}_{j,n}}  \widehat{\mathbf{{H}}}_{j,n} +  \sigma^{2}_{{h,\Delta}_{j,n}} \textrm{tr}  \left \lbrace  {\mathbf{{S}}}_{j,n} \right \rbrace \mathbf{I}_{{M}_{h}} \right \rbrace
\end{bmatrix}, \forall k,1 \leq n \leq N-1,
\end{equation*}
\begin{equation*} \label{202}
\bar{\mathbf{{D}}}_{k,n} =
\begin{bmatrix}
\mathbf{I}_{{M}_{k,N}}  \hspace*{.5mm}&\hspace*{.5mm}   \left({\widetilde{\mathbf{{S}}}}^{\frac{1}{2}}_{k,n}  \right) ^{{H}} {\widehat{\mathbf{{H}}}_{k,N}}  \vspace*{1.5mm} \\
 \widehat{\mathbf{{H}}}^{{H}}_{k,N} {{\widetilde{\mathbf{{S}}}}^{\frac{1}{2}}_{k,n}} \hspace*{.5mm}&\hspace*{.5mm} {{{\mathbf{L}}}}_n^{(\textrm{CH-HAP})} +\sum_{j=k}^{{K}}  \left \lbrace {\widehat{\mathbf{{H}}}^{{H}}_{j,N}}  {{\widetilde{\mathbf{{S}}}}_{j,n}}  \widehat{\mathbf{{H}}}_{j,N} + \sigma^{2}_{{h,\Delta}_{j,N}} \textrm{tr}  \left \lbrace  {\widetilde{\mathbf{{S}}}}_{j,n} \right \rbrace \mathbf{I}_{{M}_{h}} \right \rbrace
\end{bmatrix}, \forall k,n,
\end{equation*}
\begin{equation*}
{\widetilde{\mathbf{F}}}^{(\kappa-1)}_{k,n}= \left({\widetilde{\mathbf{D}}}^{(\kappa-1)}_{k,n} \right)^{-1}\widetilde{\mathbf{A}}_{k,n} \left(\widetilde{\mathbf{A}}^{{H}}_{k,n} \left({\widetilde{\mathbf{D}}^{(\kappa-1)}_{k,n}}\right)^{-1} \widetilde{\mathbf{A}}_{k,n}\right)^{-1} \widetilde{\mathbf{A}}^{{H}}_{k,n} \left({\widetilde{\mathbf{D}}^{(\kappa-1)}_{k,n}}\right)^{-1}, \hspace{4pt} \forall k,1 \leq n \leq N-1,
\end{equation*}
\begin{equation*}
\bar{\mathbf{F}}^{(\kappa-1)}_{k,n}= \left({\bar{\mathbf{D}}^{(\kappa-1)}_{k,n}}\right)^{-1}\bar{\mathbf{A}}_{k,n} \left(\bar{\mathbf{A}}^{{H}}_{k,n} \left({\bar{\mathbf{D}}^{(\kappa-1)}_{k,n}}\right)^{-1} \bar{\mathbf{A}}_{k,n}\right)^{-1} \bar{\mathbf{A}}^{{H}}_{k,n} \left({\bar{\mathbf{D}}^{(\kappa-1)}_{k,n}}\right)^{-1}, \hspace{4pt} \forall k,n,
\end{equation*}
\begin{align*}   \label{appb1}
\widetilde{\mathbf{T}}^{(\kappa-1)}_{k,n} =& - \textrm{log}_2 \textrm{det} \left(\widetilde{\mathbf{A}}^{{H}}_{k,n}  \left(\widetilde{\mathbf{D}}^{(\kappa-1)}_{k,n}\right)^{-1}\widetilde{\mathbf{A}}_{k,n} \right) - \textrm{tr} \left \lbrace \widetilde{\mathbf{F}}^{(\kappa-1)}_{k,n} \widetilde{\mathbf{D}}^{(\kappa-1)}_{k,n}\right \rbrace + \textrm{tr} \left \lbrace \left(\widetilde{\mathbf{F}}^{(\kappa-1)}_{k,n}\right)_{11} \right\rbrace \\ &+ \textrm{tr} \left \lbrace\left(\widetilde{\mathbf{F}}^{(\kappa-1)}_{k,n}\right)_{22} {{\mathbf{L}}}_n^{(\textrm{U-HAP})} \right\rbrace, \hspace{4pt} \forall k, 1 \leq n \leq N-1,
\end{align*}
\begin{align*}
\bar{\mathbf{T}}^{(\kappa-1)}_{k,n} =& - \textrm{log}_2 \textrm{det} \left(\bar{\mathbf{A}}^{{H}}_{k,n} (\bar{\mathbf{D}}^{(\kappa-1)}_{k,n})^{-1}\bar{\mathbf{A}}_{k,n} \right) - \textrm{tr} \left \lbrace \bar{\mathbf{F}}^{(\kappa-1)}_{k,n} \bar{\mathbf{D}}^{(\kappa-1)}_{k,n}\right \rbrace + \textrm{tr} \left\lbrace\left(\bar{\mathbf{F}}^{(\kappa-1)}_{k,n}\right)_{11} \right\rbrace \\ &+ \textrm{tr} \left\lbrace \left(\bar{\mathbf{F}}^{(\kappa-1)}_{k,n} \right)_{22} {{\mathbf{L}}}_n^{(\textrm{CH-HAP})} \right\rbrace, \hspace{4pt} \forall k,n,
\end{align*}
\begin{equation*} \label{appb2}
\widetilde{\mathbf{v}}^{(\kappa-1)}_{k,n}=\textrm{vec}\left( \widehat{\mathbf{{H}}}_{k,n} \left(\widetilde{\mathbf{F}}^{(\kappa-1)}_{k,n}\right)^{{H}}_{12} \right), \hspace{1pt} \forall k,1 \leq n \leq N-1,
\hspace{10pt}
\bar{\mathbf{v}}^{(\kappa-1)}_{k,n}=\textrm{vec}\left( \widehat{\mathbf{{H}}}_{k,N} \left(\bar{\mathbf{F}}^{(\kappa-1)}_{k,n}\right)^{{H}}_{12} \right), \hspace{1pt} \forall k,n,
\end{equation*}
\begin{equation*} \label{appb3}
\widetilde{\mathbf{\Upsilon}}^{(\kappa-1)}_{j,n}=\mathbf{I}_{{M}_{j,n}}  \otimes \left( \widehat{\mathbf{{H}}}_{j,n} \left(\widetilde{\mathbf{F}}^{(\kappa-1)}_{k,n}\right)_{22} \widehat{\mathbf{{H}}}^{{H}}_{j,n} + \sigma^{2}_{{h,\Delta}_{j,n}} \textrm{tr} \left\lbrace { (\mathbf{F}_{k,n})}_{22} \right\rbrace  \mathbf{I}_{{M}_{j,n}} \right), \hspace{4pt} \forall k,1 \leq n \leq N-1,
\end{equation*}
\begin{equation*}
\bar{\mathbf{\Upsilon}}^{(\kappa-1)}_{j,n}=\mathbf{I}_{{M}_{j,N}}  \otimes \left( \widehat{\mathbf{{H}}}_{j,N} \left(\bar{\mathbf{F}}^{(\kappa-1)}_{k,n}\right)_{22} \widehat{\mathbf{{H}}}^{{H}}_{j,N} + \sigma^{2}_{{h,\Delta}_{j,N}} \textrm{tr} \left\lbrace { (\mathbf{F}_{k,n})}_{22} \right\rbrace  \mathbf{I}_{{M}_{j,N}} \right),  \hspace{4pt} \forall k,n.
\end{equation*}
\section{A choice of ${\xi}_{k,n}$ in Subsection \ref{det}} \label{app4}
The parameter ${\xi}_{k,n}$ can be selected in such a way that it satisfies the following condition:
\begin{equation}
{\nabla}^{2}_{{\mathbf{x}}_0} {E}_{k,n} \left( {\mathbf{x}}_0 \right) + {\xi}_{k,n} \mathbf{I}_{{{M}_h}} \succeq \mathbf{0}, \hspace{6pt} \forall k,n,
\end{equation}
where ${\nabla}^{2}_{{\mathbf{x}}_0} {E}_{k,n} \left( {\mathbf{x}}_0 \right)$ is straightforwardly calculated as
\begin{equation}
{\nabla}^{2}_{{\mathbf{x}}_{0}} {E}_{k,n} \left( {\mathbf{x}}_{0} \right)  =\varrho_{k,n}  \mathbf{B}_{k,n} + \varpi_{k,n}   \mathbf{B}_{k,n} \mathbf{x}_{0} \mathbf{x}^{{H}}_{0} \mathbf{B}_{k,n}, \hspace{4pt} \forall k,n,
\end{equation}
with
\begin{equation}
\varrho_{k,n} = 2\tau_2 \hspace{1pt} \textrm{exp} \left( {c} \right) \hspace{1pt}  \textrm{exp} \left( {a} \hspace{1pt} {\left( \textrm{log} \left(   \widetilde{P}_{k,n}   \right)  \right)}^2  \right) \widetilde{P}_{k,n}^{  {b}-1} \left(  2 {a} \textrm{log} \left(  \widetilde{P}_{k,n}    \right) + {b} \right),
\end{equation}
\begin{equation}
\varpi_{k,n}   = 4\tau_2 \textrm{exp} \left( {c} \right)  \textrm{exp} \left( {a}  {\left( \textrm{log} \left(  \widetilde{P}_{k,n}   \right)  \right)}^2  \right) \widetilde{P}^{ {b}-2}_{k,n} \left( 4 {{a}}^2 \left( \textrm{log} \left(   \widetilde{P}_{k,n}   \right) \right)^2  +\textrm{log} \left(  \widetilde{P}_{k,n}   \right) \left( 4{{a}}{{b}} -2{{a}} \right) + {{b}}^2 -{{b}} + 2 {a} \right).
\end{equation}
 As ${a}<0, {b} <0,\mathbf{B}_{k,n}\succeq \mathbf{0},$ and $\mathbf{B}_{k,n} \mathbf{x}_{0} \mathbf{x}^{{H}}_{0} \mathbf{B}_{k,n}\succeq \mathbf{0}$, it suffices to choose $\xi_{k,n}$ such that
\begin{equation} \label{f0}
\xi_{k,n} \mathbf{I}_{{{M}_h}} \succeq - \widetilde{\varrho}_{k,n} \mathbf{B}_{k,n} - \widetilde{\varpi}_{k,n}   \mathbf{B}_{k,n} \mathbf{x}_{0} \mathbf{x}^{{H}}_{0} \mathbf{B}_{k,n},
\end{equation}
where
\begin{equation} \label{f01}
\widetilde{\varrho}_{k,n} = 2\tau_2 \hspace{1pt} {b} \hspace{2pt} \textrm{exp} \left( {c} \right) \hspace{1pt}  \textrm{exp} \left( {a} \hspace{1pt} {\left( \textrm{log} \left(   \widetilde{P}_{k,n}   \right)  \right)}^2  \right) \widetilde{P}^{  {b}-1}_{k,n},
\end{equation}
\begin{equation} \label{f02}
\widetilde{\varpi}_{k,n} = 4\tau_2 \textrm{exp} \left( {c} \right)  \textrm{exp} \left( {a}  {\left( \textrm{log} \left(  \widetilde{P}_{k,n}  \right)  \right)}^2  \right) \widetilde{P}^{  {b}-2}_{k,n} \left( \textrm{log} \left(   \widetilde{P}_{k,n}   \right) \left( 4{{a}}{{b}} -2{{a}} \right) + 2 {a} \right).
\end{equation}
Then, using ${\textrm{C}}^{\textrm{det}}_{2}$, \eqref{f0}, \eqref{f01}, \eqref{f02} and considering the fact that $\mathbf{x}^{{H}}_{0} \mathbf{B}^{2}_{k,n} \mathbf{x}_{0} \leq \|{\mathbf{x}}_{0}\|^2_2 \lambda_{\textrm{max}} \left(\mathbf{B}^{2}_{k,n} \right)$ as well as $ \|{\mathbf{x}}_{0}\|^2_2 \lambda_{\textrm{min}} \left(\mathbf{B}_{k,n} \right) \leq \widetilde{P}_{k,n}=\mathbf{x}^{{H}}_{0} \mathbf{B}_{k,n} \mathbf{x}_{0} \leq \|{\mathbf{x}}_{0}\|^2_2 \lambda_{\textrm{max}} \left(\mathbf{B}_{k,n} \right)$, the Parameter ${\xi}_{k,n} > {\xi}^{t}_{k,n} $ can be selected as
\begin{align*}
{\xi}^{t}_{k,n}=&-  2\tau_2 \hspace{1pt}  \textrm{exp} \left( {c} \right) \hspace{1pt}  \textrm{exp} \Big( {a} \hspace{1pt} {\left( \textrm{log} \left(  p_0   \lambda_{\textrm{max}} \left(\mathbf{B}_{k,n} \right)   \right) \right)}^2  \Big) \big(  p_0   \lambda_{\textrm{min}} \left(\mathbf{B}_{k,n} \right) \big)^{b-1}
\Big(
{b} \hspace{2pt}  p_0   \lambda_{\textrm{min}} \left(\mathbf{B}_{k,n} \right)  \lambda_{\textrm{max}} \left(\mathbf{B}_{k,n} \right)
\\ \nonumber &
+ 2\left( \textrm{log} \left(  p_0   \lambda_{\textrm{min}} \left(\mathbf{B}_{k,n} \right)   \right) \left( 4{{a}}{{b}} -2{{a}} \right) + 2 {a} \right) p_0   \lambda_{\textrm{max}} \left(\mathbf{B}^{2}_{k,n} \right)
\Big), \hspace{4pt} \forall k,n,
\end{align*}
where, we should remark on the fact that $\lambda_{\textrm{min}} \left(\mathbf{B}_{k,n} \right) = \lambda_{\textrm{min}} \left(  \widehat{\mathbf{H}}^{{H}}_{k,n}  {\widehat{\mathbf{H}}_{k,n} }   \right) +   \sigma^{2}_{{h,\Delta}_{k,n}} {{M}_{k,n}} >0, \hspace{1pt} \forall k,n$.
\bibliographystyle{IEEETran}
\bibliography{myreff1}

\end{document}